\documentclass{statsoc}

\usepackage[a4paper]{geometry}
\usepackage{graphicx}
\usepackage[textwidth=8em,textsize=small]{todonotes}
\usepackage[utf8]{inputenc}
\usepackage{amsmath, amssymb}
\usepackage{natbib}
\usepackage{hyperref}
\usepackage{multirow}
\usepackage{natbib}

\usepackage{float}
\usepackage[caption = false]{subfig}
\definecolor{verde}{rgb}{0.2, 0.7, 0.2}
\usepackage{etoolbox}

\makeatletter
\patchcmd{\@makecaption}
  {\parbox}
  {\advance\@tempdima-\fontdimen2} % decrease the width!
  {}{}
\makeatother 

\title[Density-based clustering of weighted networks]{Density-based clustering of social networks}
\author[Menardi, G. and De Stefano, D.]{Giovanna Menardi}
\address{Department of Statistical Sciences, University of Padova, Italy %\\
}
\email{menardi@stat.unipd.it}
\author[Menardi, G. and De Stefano, D.]{Domenico De Stefano}
\address{Department of Political and Social Sciences,
University of Trieste, Italy
}
\email{ddestefano@units.it}
\begin{document}

\begin{abstract}
The idea underlying the modal formulation of density-based clustering is to associate groups with the regions around the modes of the probability density function underlying the data. This correspondence between clusters and dense regions in the sample  space is here exploited to discuss an extension of this approach to the analysis of social networks. Such extension seems particularly appealing: conceptually, the notion of high-density cluster fits well the one of community in a network, regarded to as a collection of individuals with dense local ties in its neighbourhood. 
The lack of a probabilistic notion of density in networks is turned into a major strength of the proposed method, where node-wise measures that quantify the role and position of actors may be used to derive different community configurations. 
The approach allows for the identification of a hierarchical structure of clusters, which may catch different degrees of resolution of the clustering structure. This feature well fits the nature of social networks, disentangling a different involvement of individuals in social aggregations.
\end{abstract}
%Additionally, modal clustering often resorts to graph theory for the operational detection of clusters, another condition that seems particularly appropriate to deal with relational data. 
%Finally, unlike optimization methods which suffer from the resolution limit, as the popular Louvain algorithm, we are able to find small communities even in huge networks and jointly account for larger levels of aggregation.

\section{Introduction}
\subsection{Background and motivation}

Within large social communities, individuals sparsely interact with each others and usually set a tight relationship
with a limited number of subjects. Interactions favour individuals to aggregate into groups, where the 
relationships are stronger and the information flow is more intense than outside. 

The generating mechanism of these groups, albeit pervasive, is complex and often difficult to be disclosed. 
On one hand, different kinds of relationship may be established, from friendship to professional collaboration, each of them possibly with different levels of intensity. 
On the other hand, aggregation may be driven by diverse, sometimes unobserved, social mechanisms -- homophily, popularity,
ranking or influence. Depending on the context, cohesive communities may be formed, where even relationships connect each actor with most of other actors. This configuration characterizes, for instance, individual interactions, communication system, sport and team relationships \citep{carron2000cohesion}.
A different dynamic arises when one or few influential actors drive the aggregation and shape the whole organization of the community \citep{Ghalmane2019}.   
Examples of this latter behaviour are opinion or news spreading in online communities where followers are
attached to influencers \citep[e.g.][]{Wang2017}; epidemic diffusion where few prominent actors govern the outbreak \citep{medo2009}, scientific collaborations and citations where communities develop around the so-called star scientists \citep{DeStefano_etal:13}.
Here, the nature of leadership may be associated to various roles which actors carve out within the groups,
acting for instance as hubs or brokers. % \citep{Handcock}.

In this context, Social network analysis (SNA) exploits the framework offered by graph theory to translate these ideas into operational tools: any community is suitably described by a graph where nodes represent the actors and the links between them their interactions, possibly of different strength.
A wide range of methods, among which centrality or equivalence measures are just simple examples, have been spawned to express notions of social role and position. 
A standard accounts is \citet{Wasserman_Faust:94}.%, and \citet{Scott:12}. 

%\textcolor{verde}{The identification of groups in networks is usually referred to as \emph{community detection} or \emph{graph clustering}. The two notions are often acknowledged to have a slight different meaning; in fact, there is no clear agreement about their specificities, hence we will use the two terms exchangeably. See \citet{Abbe17} for a thorough discussion. 
%Disregarding nomenclature, 
While the underlying scope to find groups in network may follow different routes, these are usually defined as locally densely connected set of nodes. The correspondence between groups of subjects and their inner connection density, as well as the possible role of influential individuals within communities, suggest us to extend the ideas underlying the density-based approach for clustering nonrelational data to the network framework.
The \emph{modal} formulation of this approach associates clusters with the domains of attraction of the modes of the density function underlying the observed data, namely clusters correspond to dense regions of the sample space.
While network data unarguably prevent the definition of a probabilistic notion of a density function defined on the nodes,  
the two notions of group are in agreement conceptually. Operationally,
modal clustering often resorts to graph theory to detect clusters, which further favours
the extension of this formulation to network data. As a fortunate side effect the modal approach
allows for the identification of a hierarchical structure of clusters, which may catch
different degrees of resolution of the clustering structure.

Based on these ideas, the aim of this work is to discuss a method to find clusters of nodes within a network structure, while accounting for relationships of different strength. Consistently with the cluster notion shared by the nonrelational density-based approach, we focus on aggregation mechanisms driven by the attraction exerted by influential actors, on the basis of different ``leadership'' roles as detected by means of alternative node-wise measures. Note that this perspective is largely neglected by the inherent literature, most focusing on the concept of mutual cohesiveness within communities.
%\textcolor{red}{We also discuss how the proposed approach adopts a quite different perspective with respect to the existing community detection methods allowing for the identification of clusters determined by rather diverse attachment mechanisms (e.g., leadership based communities) whereas in the existing literature, usually the cluster formation rationale is mainly based on the concept of mutual cohesiveness which can be not suitable for a wide range of social phenomena.}

The  paper is organised as follows. After a brief review of clustering approaches for networks, we overview the modal clustering formulation in metric spaces. Then, we discuss its extension to network data, in both, the unweighted and weighted network framework.
The procedure is illustrated on some simple archetypal networks characterized by different community configurations, on a number of benchmark examples with a known community structure, and on a comprehensively complex original dataset to identify groups of researchers within the community of the Italian academic statisticians. A discussion concludes the paper.

\subsection{Overview of the related literature}

\emph{Community detection} refers to the general aim of partitioning networks in subsets of nodes, which share some common properties and play similar roles in the relational structure. Similarly to the nonrelational framework, this task is, in fact, far from being accurately defined. Thus, while the general purpose usually translates into the task of identifying assortative groups with dense inner connections, a different perspective would also include the search of disassortative structures with weaker interactions within, rather than between communities. 

The lack of a precise definition of cluster, along with the unsupervised nature of the problem, have led on one hand to the proliferation of a voluminous amount of literature on this topic and, on the other hand, to confusing taxonomies of methods designed for the scope. A lack of a consistent terminology has determined expressions as \emph{network} or \emph{graph clustering}, \emph{module}, \emph{block} or \emph{community detection} to be either used interchangeably, or carry slightly different, yet ambiguous, connotations. In this confounding panorama, methods are easier classified on the basis of their technical details and algorithmic implementations \citep[e.g.,][]{FORTUNATO2010, azaouzi2019community}, which yet disguises the more relevant notion of cluster underlying them.
Reviewing all these methods is then an awkward task which we cannot engage without crossing over the scope of the paper. 
%\textcolor{red}{Here, we limit to set some boundaries by providing a coarse overview of the main different goals and motivations for finding groups in networks, with a slightly more detailed focus on the methods closer to the perspective we follow in our contribution. (Forse questo puo pezzo puo andare solo nella risposta ai referee)}
For our purpose, we limit to set some boundaries by providing a coarse overview of the main different goals and motivations for finding groups in networks, and refer back to the insightful review of \citet{rosvall2019different}, where the reader will find further details and references.  
%At the same time, we follow the coward route to use the terms cluster, community, \emph{et similia} excheangeably in the rest of the paper. 
At the same time, we use the terms cluster, community, groups and so on exchangeably in the rest of the paper. 

The first, perhaps most widespread approach to find clusters in networks aims at identifying densely interconnected nodes compared to the other nodes. Due to the generality of this principle, methods differ in the way it is translated into operational implementations. Several formulations rely on detection of actors or edges with high centrality, as for instance, the very popular method of Girvan-Newman \citep[GN,][]{NewmanGirvan}, a divisive algorithm for undirected and unweighted graphs based on edge-betweenness, afterwards generalized by \citet{Chen2006}. 
Further methods relying on a similar ground build on the optimisation of the cluster modularity \citep{Danon2005}, so that each community will include a larger number of inner edges than expected by chance. The Louvain method is unarguably one of the most popular representative of this category \citep{Blondel_2008}. 
The aforementioned methods result in cohesive communities where transitivity is high and each actor is highly connected to each other inside the group. Notwithstanding, the idea of high density within a group may be also intended as the one arising in star-shaped clusters, where density is concentrated in the figure of some hubs attracting less prominent actors. Evidence of such a theoretical mechanism of aggregation has been explained by \citet{goyal2006economics} as a combination of small-world behavior guided by  the presence of interlinked stars.
In fact, this principle has been largely neglected by SNA, with the works of \citet{kloster2014heat}, based on the local optimization of the so-called conductance and, to some extent, \citet{falkowski2007dengraph} representing an exception. This is also the route we follow. 

A further facet of the clustering problem in networks, known as \emph{cut-based} perspective, aims at partitioning networks in a fixed number of balanced groups with a small number of edges between them, and no guarantees about a possible denser structure of inner connection. In this context, networks are often of a mesh- or grid-like form. Methods in this class refer back to the seminal work of \citet{kernighan1970efficient} and often build on the spectrum of of the data. Examples are \citet{Pothen, wang2017weighted}. 

%The block-modeling approach follows a completely different purpose, relying on the fundamental concept of \emph{structural equivalence}. 
	
The block-modeling approach follows a completely different purpose, relying on the fundamental concept of node equivalence, of which \emph{structural equivalence} is the most used. Disregarding the similarity of nodes, groups are here based on more general patterns that include disassortative communities and include nodes that serve, within the network, a similar structural role in terms of their connectivity profile. A first formalization in terms of non-stochastic blocks can be found in \citet{LorrainWhite}, while \citet{holland1983} gave rise to the stochastic counterpart, later generalized to the weighted framework \citep{Aicher2014} and largely applied in various contexts See \citet{lee2019review} for a recent review.

\section{Clusters as dense sets}
\subsection{Modal clustering of non-relational data}\label{sec:modal}

Modal clustering delineates a class of methods for grouping non-relational data defined on a metric, continuous space, and building
on the concept of clusters as ``regions of high density separated from other such regions by regions of low density'' \citep[p. 205]{Hartigan:75} 
Formally, the observed data $(x_{1}, \ldots, x_{n})'$, $x_i\in \mathbb{R}^d$, $i=1, \ldots, n$, are supposed to be 
a sample from a random vector with (unknown) probability density function $f$. The modes of $f$ are regarded to as the archetypes of the clusters, which are in turn represented by their domain of attraction.

\begin{figure}[t]
\includegraphics[width=0.32\textwidth]{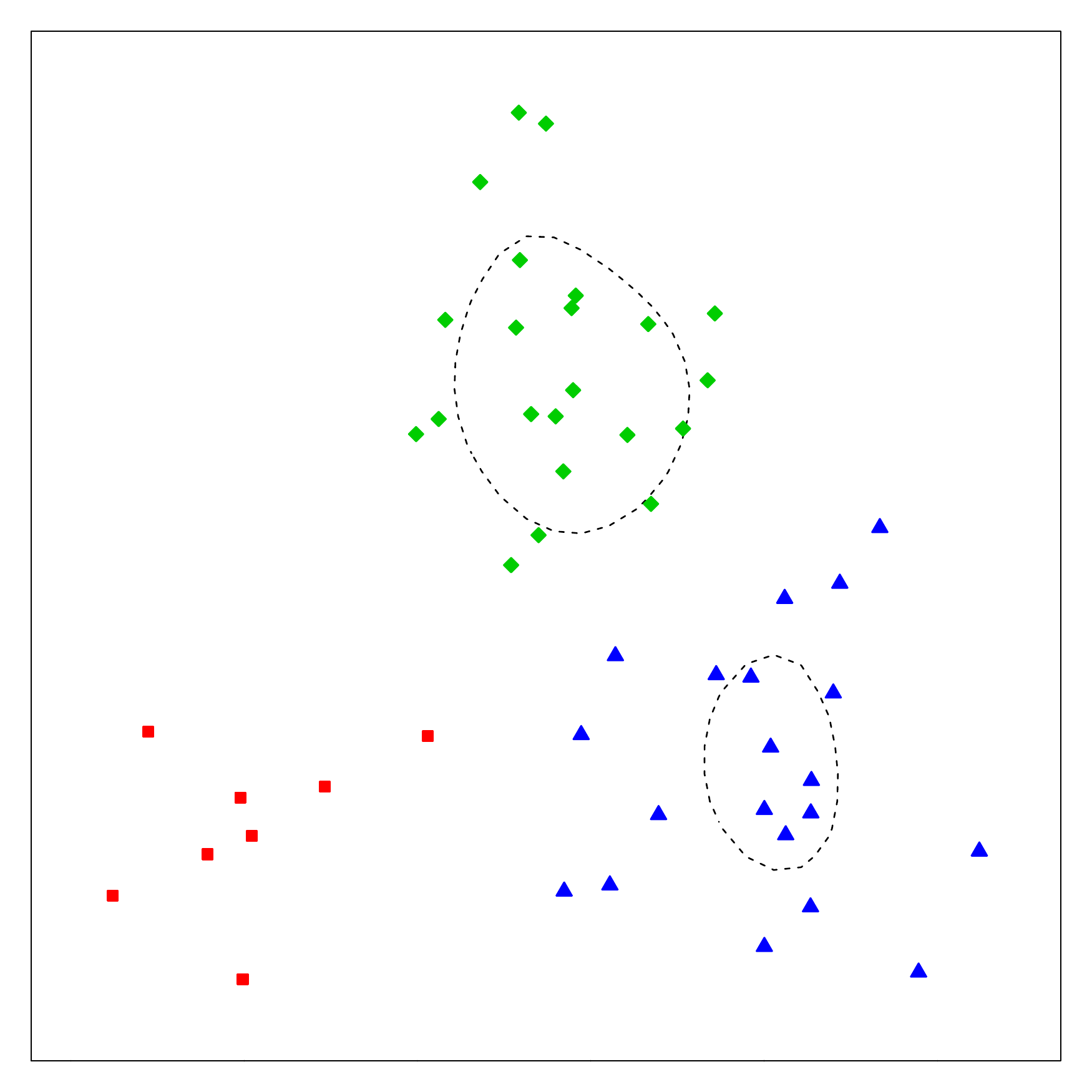}\hspace{-.5cm}
\includegraphics[width=0.38\textwidth]{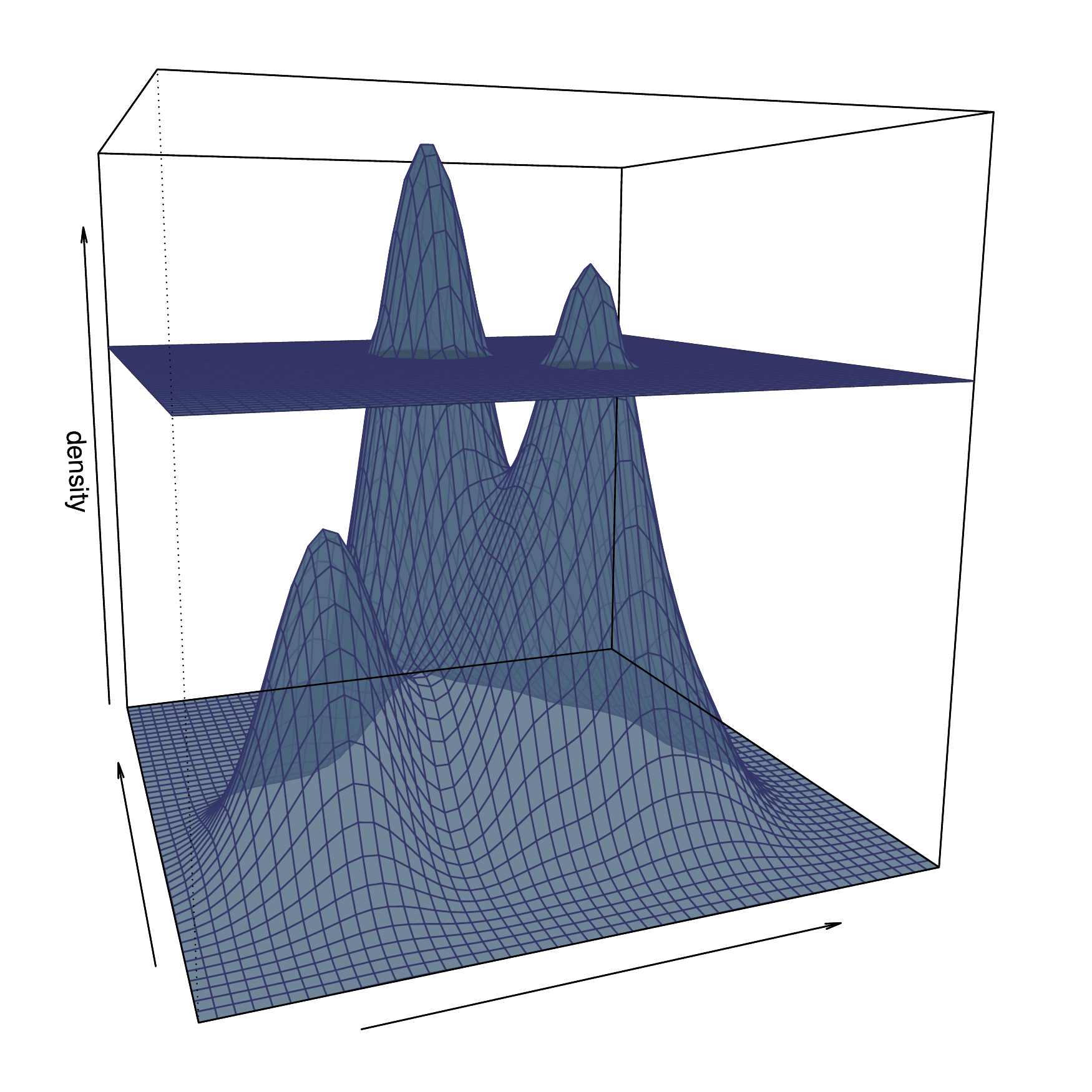}\hspace{-.4cm}
\includegraphics[width=0.3\textwidth]{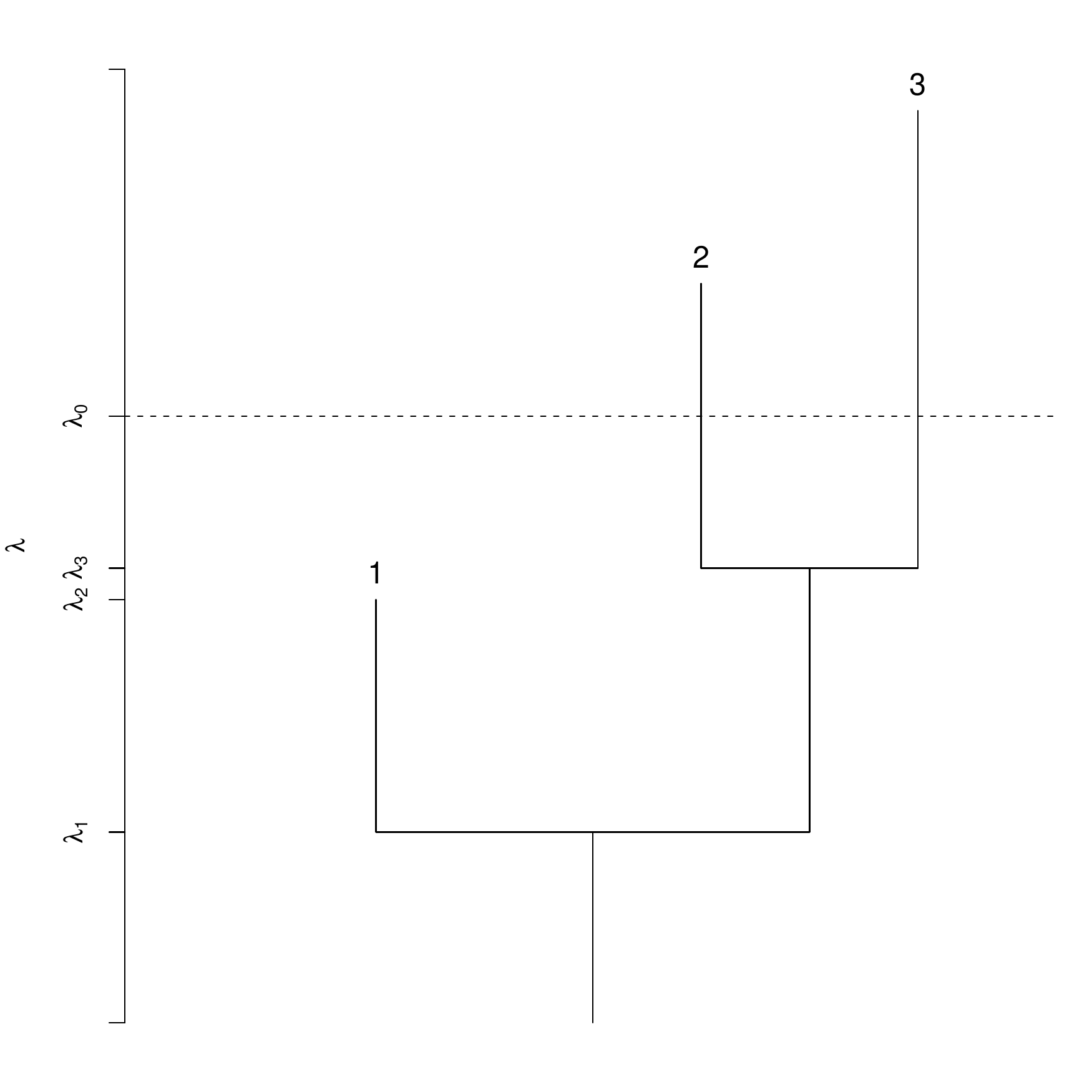}
\caption{A sample from three subpopulations and the associated contour set at a level $\lambda_0$ (left). 
The threshold $\lambda_0$ defines a section of the trimodal underlying density function (center) and identifies
two connected regions. On the right, the cluster tree indicates the number of connected components
for varying $\lambda$ and the total number of clusters, corresponding to the leaves.}\label{fig:modal}
\end{figure}

The practical identification of the modal regions may occur according to different directions. 
One of them associates the clusters to disconnected density level sets of the sample space, without attempting explicitly the difficult task of mode detection.
The key idea is that, when there is no clustering structure, $f$ is unimodal, and any section of $f$, at a given level $\lambda$, singles out a connected (upper) level set:
%\begin{equation}\label{eq:level_set}
$L(\lambda) = \{x\in \mathbb{R}^d: f(x)\geq \lambda\}$.
%\end{equation}
Conversely, when $f$ is multimodal, $L(\lambda)$ may be either connected or disconnected, depending on $\lambda$. 
In the latter case, it is formed by a number of connected components, each of them associated with a region of the sample space including at least one mode of $f$.
Since a single section of $f$ could not reveal all the modes of $f$, $\lambda$ is moved along its feasible range, giving rise to a hierarchical structure, known as the \emph{cluster tree}, which provides the number of connected components for each $\lambda$. Each leaf of the tree describes a \emph{cluster core}, defined as the largest connected component of the density level sets which includes one mode.
Figure \ref{fig:modal} illustrates a simple example of these ideas: cluster cores associated with the two highest modes are identified by the smallest $\lambda$ larger than $\lambda_3$, while the smallest $\lambda$ larger than $\lambda_1$ identifies two connected components whose one is the cluster core associated to the lowest mode.

Note that while the cluster tree resembles a dendrogram, the whole procedure cannot be included in the class of hierarchical techniques. 
These explore, within the same run, all the partitions with a number of clusters ranging from 
one to $n, $ by subsequent splits (divisive algorithms) or aggregations (agglomerative algorithms). Conversely, in the cluster tree, 
the leaves are themsevels veritable clusters, instead of single observations, and their number is then an estimate of the number of clusters. 
Hence, with respect to a dendrogram, the cluster tree enjoys a different, more insightful interpretation.
The height of the leaves corresponds to the density level at which the associated mode appears,
thus providing an indication of the cluster prominence. Finally, the hierarchical structure of the tree
allows for catching possible different degrees of resolution of the clustering.
In the example illustrated in Figure \ref{fig:modal} the number of modes is three, but
the two highest ones pertain to the same macro-group, at a lower
level of resolution, hence the leaves associated to the two groups collapse to a single branch accordingly.

As the union of the cluster cores is not a partition of the sample space, unallocated points are assigned to the cluster
cores according to a supervised scheme of classification, generally accounting for their density.

Operationally, clustering involves two main choices: first, a density estimator
is required and this is typically selected among the nonparametric methods. 
Second, for each examined threshold $\lambda$ it is to establish whether the associated level
set is connected and what are its components. Since there is no obvious method to identify connected sets in a multidimensional space, graph theory comes to this aid. A graph is built on the sample points and the connected components 
of the subgraphs induced by the level sets are then easily detected. The reader is referred to \citet{Menardi:15} for further details about modal clustering.

\subsection{Modal clustering of social networks}
\subsubsection{Defining density on networks}\label{sec:density}

For the current formulation, we regard to social networks as undirected graphs
$\mathcal{G} =\{\mathcal V, \mathcal E\}$ consisting of a set $\mathcal V =\{v_{1}, \dots, v_{n}\}$
of nodes -- the actors of the network-- and a set $\mathcal E=\{e_{ij}\}$ of $m$ links or edges, $i \neq j = 1, \ldots, n,$
representing relations between pairs of nodes. 
Depending on the nature of the observed relationships, the elements of $\mathcal E$ assume different forms: 
in binary networks the $e_{ij}$ will take values in $\{0, 1\}$, denoting the absence and the presence of a link, respectively, while real nonnegative values of $e_{ij}$ will account for different strengths of the relationship in weighted networks.
In order to represent a given network $\mathcal{G}$ it is possible to define a $n \times n$ adjacency matrix $\mathbf{A}$ whose elements $a_{ij} = e_{ij} $.

The notion of high-density regions highlighted in the previous section suggests a natural counterpart in network analysis,
where clusters are often referred to as sets of actors with dense relationship patterns \citep[see, among others, ][]{Moody:01}.
However, network objects are subject to an inherent limitation, as their properties can be established in geodesic 
terms only. In particular, a probabilistic notion of density cannot be defined  
and shall be intended in a less formal way, reflecting some intuitive meaning of 
cohesiveness. 

We are naturally tempted to borrow the concept of density and akin notions from graph theory.
The density of a subgraph $\mathcal{H} \subseteq \mathcal{G}$ is defined as the proportion of 
all possible edges of $\mathcal{H}$ which are actually observed. 
In fact, density definition as a node-wise measure is arbitrary 
as a subgraph $\mathcal{H}_v$ is required to be associated to each node $v$. 
For instance, one could set $\mathcal{H}_v = \{\mathcal V_v, \mathcal E_v\}$ as the subgraph
having the nearest neighbours of $v$ as nodes,
or focus on the single node $\mathcal V_v = v$ and its incident edges $\mathcal E_v$
thus recasting to the notion of (possible weighted) degree. 
In fact, consistently with the previous one, a wider set of candidates to quantify local density is represented by measures of 
connectivity or measures of centrality, which evaluate, somehow, the role as well as the 
prominence of each actor in a network. 
It is worthwhile to observe that the choice of a node-wise density measure is not 
inconsequential with regard to the subsequent interpretation of clusters, and different choices would entail a different concept of cluster. 
For example, the notion of degree accounts for the rate of the activity of individual nodes in the network, so that
high-degree actors act as ``hubs'' and play a central role in the overall connectivity. 
Alternatively, by measuring the proportion of times a node works as a broker connecting nodes otherwise disconnected in the network, betweenness evaluates the influence of the actors with respect to the information flow in the network.
%Closeness, on the other hand, accounts for the reachability of the actors in the network, as measured by evaluating the shortest paths connecting each actor to all others \citep{Freeman}.
Despite in the following we adopt well-known node centrality measures only, any function defined on the node set $\mathcal{V}$ or alternative node-wise measures that allow to quantify the role and/or position of each node in the network can be used. This allows our procedure to be more flexible than other methods based on optimisation of a given node (or edge) function. 

While, in general, the above mentioned measures do not sum up to one, as it would be required by a density function,
they can be easily normalised to this purpose, but for the subsequent developments this is not strictly necessary.

%%%%%%%%%%%%%%%%%%%%%%%%%%%%%%%%%%%%%%%%%%%%%%%%%%%%%%%%%%%%%%%%%%%%%%%%%%%%%%%%%%%%%%%%
%clustering monoplex unweighted%%%%%%%%%%%%%%%%%%%%%%%%%%%%%%%%%%%%%%%%%%%%%%%%%%%%%%%%%%
%%%%%%%%%%%%%%%%%%%%%%%%%%%%%%%%%%%%%%%%%%%%%%%%%%%%%%%%%%%%%%%%%%%%%%%%%%%%%%%%%%%%%%%%

\subsubsection{Clustering of unweighted networks}\label{sec:clustering_monoplex}

Consider a binary network $\mathcal{G} = \{\mathcal{V}, \mathcal{E}\}$, where $\mathcal{E} = \{e_{ij}\}$
and $e_{ij} \in \{0, 1\}.$ To perform clustering, we select a node-wise measure of density 
$\delta: \mathcal{V} \mapsto \mathrm{R}^{+} \cup \{0\}$ as discussed in the previous section. 
Afterwards, we may proceed to cluster the nodes according to the modal formulation 
illustrated in Section \ref{sec:modal}, i.e. actors are clustered together when they have density above the examined threshold and they are connected.
With respect to the nonrelational framework above, we further benefit of the fact that the connected components of the high-density level sets may be identified as the connected components of the induced subgraphs, namely 
the maximal set of nodes such that each pair of nodes is connected by a path. An operational route is a represented by the following scheme: 

\begin{enumerate}
\item Compute the density of the relationships of each actor: $\delta(v_1), \dots, \delta(v_{i}),$ $\dots, \delta(v_{n})$. 
Clusters will be formed around the modal actors, namely actors with the densest relationship patterns.
\item For $0 < \lambda < \max_{i}\delta(v_{i}):$
\begin{itemize}
\item Determine the upper level set $\mathcal{V}_\lambda = \{v_{i} \in \mathcal{V}: \delta(v_{i})\geq \lambda\},$ 
\item Build the subgraph $\mathcal{G}_{\lambda} = (\mathcal{V}_\lambda, \mathcal{E}_\lambda) \subset \mathcal{G}$ where
 $\mathcal{E}_\lambda =\{e_{ij}(\lambda)\}$
 and
$$e_{ij}(\lambda) = \left \{ \begin{array} {ll}
e_{ij} &\mbox{if } (v_i, v_j\in \mathcal V_\lambda) \\
0 &\mbox{otherwise} 
\end{array} 
\right. $$
\item Find the connected components of $\mathcal{G}_{\lambda}$.  
\end{itemize}
\item Build the cluster tree by associating each 
level $\lambda$ to the number of connected components of $\mathcal{G}_{\lambda}$. 
\item Identify all the lowest $\lambda$ for which the branches of the tree represent the leaves, and form
the cluster cores as the connected components of the different associated $\mathcal{G}_\lambda$. 
 \end{enumerate}
Essentially, at each threshold $\lambda$ we evaluate the connected components of $\mathcal{G}_\lambda$, the subgraph
formed by the nodes with density above $\lambda$ and the only connections between them.  
The scheme usually leaves unallocated a number of actors with low-density patterns,
when they do not univocally pertain to a modal actor. 
Depending on the aim of clustering and on subjects-matter considerations,
part, or all of them may be either left unallocated or assigned
to the cluster for which they present the highest density $\delta(\cdot)$. %In the case of ties, the actor allocation is random among the cluster candidates. \todo{rivedere, non è proprio così}   

The described way of proceeding entails the early identification of clusters as formed by actors with the highest density, i.e. the leaders of the community, and the subsequent aggregation to the formed clusters of actors with less prominent role. In this sense, and consistently with the non-relational version of modal clustering, the final clusters are then described by the domains of attraction of the community leaders.

\subsubsection{Clustering of weighted networks}\label{sec:weighted}

Let us now consider a weighted network $\mathcal{G}= \{\mathcal V, \mathcal E\}$, where $\mathcal E =\{e_{ij}\}$ 
and $e_{ij} \in \mathbb{R}^{+} \cup \{0\}$, \emph{i.e.} the link weight is proportional to the strength of the relationship between the
two incident nodes and it is set to zero when the two nodes are not linked.  

As a first natural ploy to account for real-valued edges,
we consider density measures for weighted networks. Indeed, the
generalisation of these measures to weighted networks has been historically a somewhat controversial matter which
cannot be tackled without considering the nature of the data, the goal of the analysis, and
subject-matter knowledge. However, for most of the mentioned candidate measures $\delta$,  
there exist a reasonable weighted counterpart. The degree, for instance, is easily
extended to measure centrality in weighted networks by summing up the weights incident with each node.
This allows considering prominent an actor not only when he has many connections, but also
when the strength of these connection is large.
We refer the reader to the existing literature for a discussion
about the specification of descriptive measures for weighted networks \citep{Opsahl_etal:10}.
  
In the presence of relationships of different strengths, we need to further adjust the 
presented procedure. Indeed, a possible weak connection between two high density actors does
not appear as a sufficient condition for them to be clustered. Thus, we account for the weights
on the basis of the following simple idea: two actors are clustered together when they have density
above the examined threshold and they are \emph{strongly} connected.
Actors presenting a weak relationship with their neighbours are merged into the same cluster at a
lower level of density. Here, the strength of the connection is intended as relative to the set of connections of each node. While this is consistent with the natural idea that prominent actors exercise more influence over their strong connections and less influence over their weak connections, its implementation may take various forms. 
The following scheme provides two options of possible operational routes:
\begin{enumerate}
\item Compute the density of each actor, $\delta(v_1), \dots, \delta(v_{i}),$ $\dots, \delta(v_{n})$,
with $\delta$ an appropriate measure of node-wise density accounting for the weights of the edges;
\item For each node $v_i, i=1, \ldots,n$, identify the incident edge with maximum weight $e_{im} = \underset{j: \mbox{ }e_{ij}\in \mathcal{E}}{\max} {e_{ij}};$
\item For $0 < \lambda < \max_{i}\delta(v_{i}):$
\begin{itemize}
\item Determine the upper level set  $\mathcal{V}(\lambda) =\{v_i \in \mathcal{V}: \delta(v_i) \geq \lambda\}$
\item Build the subgraph $\mathcal{G}_\lambda = \{\mathcal V_\lambda, \mathcal E_\lambda \}$,
where $\mathcal{E}_\lambda = \{e_{ij}(\lambda )\}$  and $e_{ij}(\lambda)$ can be defined according to the two alternative options, denoted by `AND' and `OR' respectively.  
\begin{description}
\item[option AND]
$$e_{ij}(\lambda) = \left \{ \begin{array} {ll}
e_{ij} &\mbox{if } (v_i, v_j\in \mathcal V_\lambda) \cap ((e_{im} = e_{ij}) \cap (e_{jm}  = e_{ij}))\\
0 &\mbox{otherwise} 
\end{array} 
\right. $$
\item[option OR]
$$e_{ij}(\lambda) = \left \{ \begin{array} {ll}
e_{ij} &\mbox{if } (v_i, v_j\in \mathcal V_\lambda) \cap ((e_{im} = e_{ij}) \cup (e_{jm} = e_{ij}))\\
0 &\mbox{otherwise} 
\end{array} 
\right. $$

\end{description}
\item find the connected components of $\mathcal{G}_\lambda$      
\item update $e_{im} = \underset{j: \mbox{ }e_{ij}\in \mathcal{E}\setminus\mathcal{E}_\lambda}{\max} {e_{i,j}};$
\end{itemize}
\item Find the connected components of $\mathcal{G}_{\lambda}$.
\item Build the cluster tree by associating each 
level $\lambda$ to the number of connected components of $\mathcal{G}_{\lambda}$.
\item Identify all the lowest $\lambda$ for which the branches of the tree represent the leaves, and form
the cluster cores as connected components of the different associated $\mathcal{G}_\lambda$. 
\end{enumerate}

Essentially, at each $\lambda$, we identify the connected components of $\mathcal{G}_\lambda$ which are formed by the nodes with density above $\lambda$. According to ``option AND'' the additional condition for aggregation is that these nodes represent their reciprocal strongest connection among those not examined yet; conversely, according to ``option OR'' the  condition is loosen by requiring that such connection is the strongest for just one of the actors. 
The two options, albeit not exhaustive, correspond to different ways of disentangling network complexity and defining the underlying network group structure. 
With the tight AND option, aggregation is harder to occur, hence leading to a large number of highly homogeneous clusters. The resulting partition is mostly driven by the importance of the relations among nodes rather than by their relative importance within the whole network.
According to the ``OR option'', where the aggregation condition is more frequently satisfied, more parsimonious partitions are created, with clusters mostly driven by the attraction hold by the high density nodes, namely the leaders, on the lower-density ones. 
Note that this way of proceeding does not guarantee that all the weights are scanned while scanning the density values,
i.e. at the lowest considered $\lambda$, the weakest connections between some pairs of actors might not be
accounted for. Since in practice these connections are negligible as, by construction, the weakest ones,
we simply circumvent this problem by identifying, at the end of the density scanning, the connected components of the network
disregarding the weights of the connections.  

The clustering procedure eventually entails the formation of singleton clusters: suppose that three connected nodes 
$u, v,$ and $z$ have all density above a given $\lambda$, but while the strongest relationship of $u$ is with $v$, the strongest
relationship of $v$ is with $z$ and \emph{viceversa}.
Then, with the AND option, $v$ and $z$ will fall in the same cluster while $u$ will be a singleton cluster which will be aggregated to
the other at a lower $\lambda$. 

Unallocated actors are finally classified to the cluster core at which they present highest density, like in the unweighted setting. 

\section{Empirical analysis}

\subsection{Aims and implementation details}
%The current section aims to illustrate the aggregation mechanism at the basis of the proposed method for different community configurations, also with respect to the selected node-wise measure.
%We consider as density measures three alternative indexes of centrality designed to catch different roles and community configurations within a network: degree centrality evaluates the actor importance in terms of number of relationships with other members of the community; betwenness centrality, by counting the number of times actor work as bridges to connect other members, evaluates their strategic role in terms of brokerage skills; finally, local density, by shifting the attention from single actors to their nearest neighbourhood, relaxes the focus on centralized groups and identifies shared leaderships. The considered measures are consistently adjusted for use in weighted networks. 

The current section aims to illustrate the aggregation mechanism at the basis of the proposed method for different community configurations, also with respect to the selected node-wise measure.
We consider as density measures three alternative indexes of centrality designed to catch different roles and community configurations within a network: degree centrality evaluates the actor importance in terms of number of relationships with other members of the community; betwenness centrality, on the other hand, by counting the number of times actor work as bridges to connect other members, evaluates their strategic role in terms of brokerage; finally, local density, by shifting the focus from single actors to their nearest neighbourhood (i.e., nodes at geodesic distance equal to one from the focal actor), relaxes the focus on centralized groups and identifies shared leaderships. The considered measures have been consistently adjusted for their use in weighted networks.

With the aim of comparison, the considered examples are run also by considering a few competing community detection methods, mostly selected because of their wide popularity: the Girvan Newman (GN) method and its extension to weighted networks, the Louvain method and Stochastic Block Models (SBM). 

All the analyses are run in the R computing environment \citet{R} with the aid of libraries \texttt{igraph} \citep{igraph}, \texttt{sna}\citep{sna}, \texttt{sbm}\citep{sbm}. The proposed method has been implemented within the \texttt{DeCoDe} package (Density-based Community Detection), available on the author webpage\footnote{https://homes.stat.unipd.it/giovannamenardi/content/software}.

\subsection{A simple illustrative example} %\textcolor{red}{(qua metterei toy)}}

For the sake of illustration, we consider as a first example of our empirical analysis some unweighted archetypal networks where the community structure is determined by the presence of high-density nodes. 

\begin{figure}[t]
\begin{center}
\includegraphics[width=.97\textwidth]{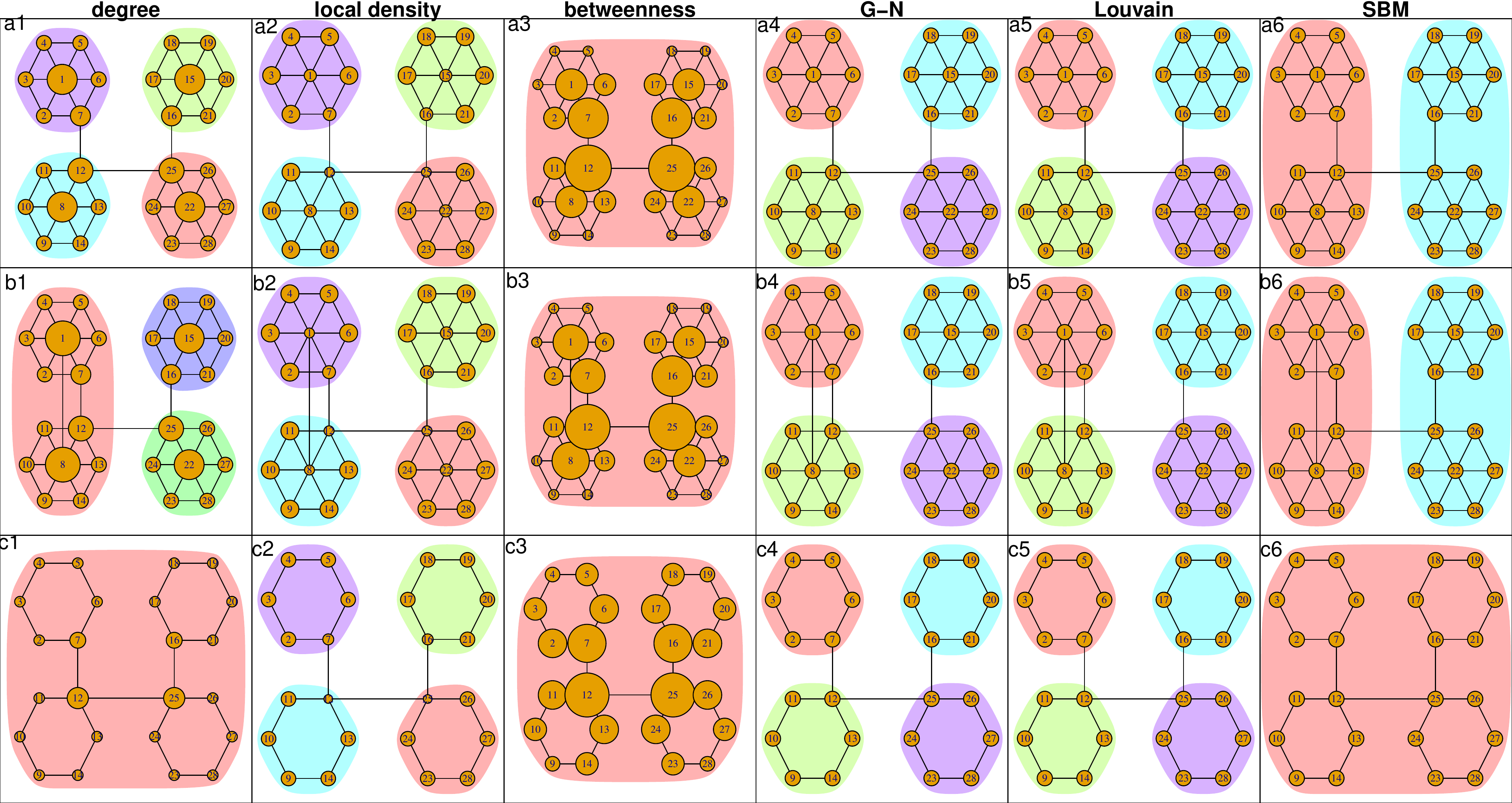}
\hspace*{.115cm}\includegraphics[width=.97\textwidth]{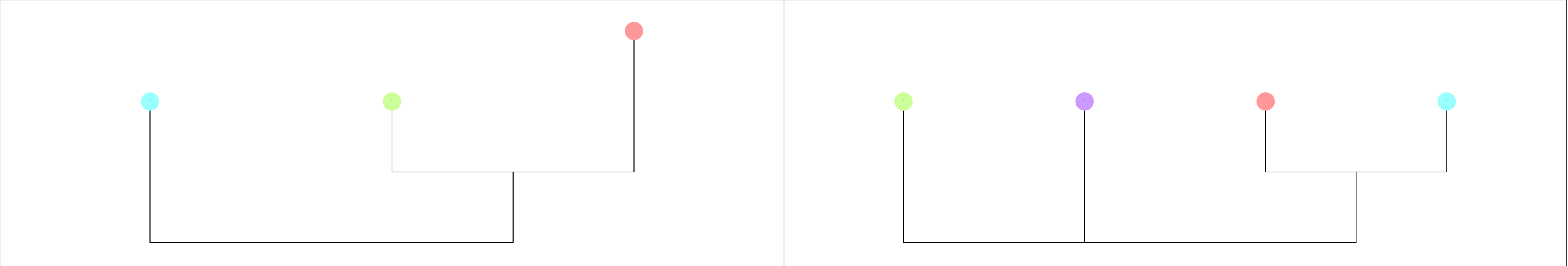}
\caption{At each row a slightly changed version of the same toy network: in the first one four hubs not linked directly; in the second row two of them are connected by a link; in the third row the hubs have been removed. At each column the clustering produced by the density-based procedure built on different measures, GN, Louvain, and SBM. Clusters are marked with different colors. In the first three columns actor size is proportional to their density. In the bottom panel, the cluster trees associated with the density-based partitions into 4 and 3 clusters.}
\label{fig:toy}
\end{center}
\end{figure}

The simple network displayed in the top row of Figure \ref{fig:toy} highlights 4 hubs standing out among 28 actors, labelled as $5, 8, 15$, and $22$. Each of the four hubs drives the information flow from and towards six actors having a less prominent role. Density-based clustering built on degree centrality reflects the hub dominance by identifying 4 clusters headed by the leaders (Figure \ref{fig:toy} a1). 
In fact, if the leaders were connected - middle panel of the Figure, where a tie links actors $1$ and $8$ - the clustering configuration would change accordingly, and a single group would be formed by all the followers of the leader dyad (Figure \ref{fig:toy} b1). 
In the lack of hubs - bottom row of Figure \ref{fig:toy}, where actors $5, 8, 15, 22$ have been removed from the network - density-based clustering built on the degree fails to identify groups (Figure \ref{fig:toy} c1), which are better identified by alternative node-wise measures accounting for a decentralized leadership. By considering, for example, the local density based on the nearest neighborhood, modal clustering detects four clusters in all the three versions of network (second column of Figure \ref{fig:toy}).
Conversely, if the the analysis focuses on the strategic role of the actors, the leadership is rather drawn by actors $12$ and $25$, acting as brokers which connect nodes otherwise disconnected in the network. With this changed aim in mind, a different structure characterizes the network, as the whole community is compact around the leaders. Consistently, density-based clustering built on betweenness detects just one cluster in all the three versions of network, leaded by the connected brokers (third column of Figure \ref{fig:toy}).   

The cluster trees provide further information by identifying the hierarchy of the communities. Thus, in the four clusters configurations the more central communities aggregate first, whereas in the three clusters configuration the first merge occurs between the largest cluster and the closest one (bottom panel of Figure \ref{fig:toy}). 

Compared to density-based clustering, GN and the Louvain method, mostly driven by the idea of modularity within a community, identify 4 clusters in all variants of the network, thus behaving like modal clustering with local density (Figure \ref{fig:toy}, fourth and fifth columns). SBM identifies an optimal partition in two clusters in the presence of hubs, and one cluster only in the absence of hubs (Figure \ref{fig:toy} a6, b6, and c6 respectively). 

% \textcolor{red}{(Proverei ad agganciare questa parte per dire che questo modo di trovare cluster ha riferimenti teorici...) The notion of cluster implied by the search of high density actors (in terms of some node-wise measure) implies a sort of attraction mechanism in which influential nodes shape the community structure. 
% Examples of such aggregations are for instance: opinion or news spreading in online communities (where followers are attached to influencers and these latters determine the form of their groups,  see for instance \citet{Wang2017}), epidemic diffusion (where few prominent actors govern the  epidemic outbreak in their respective clusters, see \citet{medo2009}), scientific collaboration and citations (where scientific communities are usually wrapped around the so-called star scientists \citet{DeStefano_etal:13}). 
%In the network literature, there are evidence of such a theoretical mechanism of aggregation, for instance when emerges a combination of small-world behavior guided by  the presence of interlinked stars \citep{goyal2006economics}.}

%\textcolor{red}{We use some unweighted archetipical networks in which the community structure is determined by the presence of high-density nodes. Let focus in particular on the case of community structures determined by the interlinked stars presence.
%As stated, for the modal clus tering approach, a suitable measure of node-wise density in this case would be the node degree.}

%As a first illustration of the proposed technique, we consider the toy network  

\subsection{Benchmark examples}
As a second step of the empirical analysis, we explore the behaviour of our method in some popular real datasets where a ground truth community membership is assigned. The choice of evaluating results in term of a true labeling, rather common in clustering, is motivated by our will of not being biased towards specific community configurations. On the other hand, it is worth highlighting that the possible identification of community structures diverse from the defined true labels would not necessarily imply a failure of the applied clustering method. Such possible result would just reflect that the true clusters have a configuration different from the one that each method is designed to detect.   

%Our aim is to compare the results in terms of employed node-wise measure on the actor set, and with respect to the performance of some other popular clustering methods. 
%\textcolor{red}{In particular, we use our density based clustering approach considering as measures of node-wise density the node degree, the betweenness, the local density and local density (DEFINITO PRIMA) possibly adjusted to account for the edge weights. Both the OR and the AND options are illustrated in the weighted networks. 
%As other methods, we again consider the Girvan Newman method, the Louvain method, and stochastic block model, whenever their application is computationally feasible and makes sense in the considered settings. Concerning the SBM, the number of communities has been selected so as to minimize the integrated classification likelihood (ICL) criterion \citep{daudin2008mixture}.}
The agreement between the true and the detected membership has been measured in terms of normalised mutual information \citep[NMI,][]{Danon2005} which increases for improved quality and associates the maximum value 1 to a perfect agreement.
%Consistently with the previous section, we consider as competing measures of node-wise density the degree, the local density and the betweenness, possibly adjusted to account for the edge weights. Both the OR and the AND options are illustrated in the weighted networks. As competitors, we again consider the Girwan Newman method, the Louvain method, and stochastic block model, whenever their application is computationally feasible and makes sense in the considered settings. 

\paragraph{Zachary Karate Club network}
The well-known Karate Club data \citep{karate} describes the network of friendships between 34 members of a karate club at a US university in the 1970s. 
The network is in principle weighted, with the strength of connections given by the number of common activities of the club members. In fact, we run the empirical analysis on both the weighted network and on its binary version, built by neglecting the strength of connections.
Due to a dispute between the administrator `John A' and the instructor `Mr Hi', the club split into two factions, here representing the benchmark membership. The two factions are then built around the leadership of John A and Mr Hi, which play a special role in terms of both direct influence on the club members, and influence on the information flow to and from the actors. %In this sense, centrality measures such as betwenness and, especially, degree, seem to to be the most appropriate to recover the faction membership via density-based clustering.    
 
\begin{figure}[t]
%\vspace{-1.7cm}
\centering
\includegraphics[scale=.2]{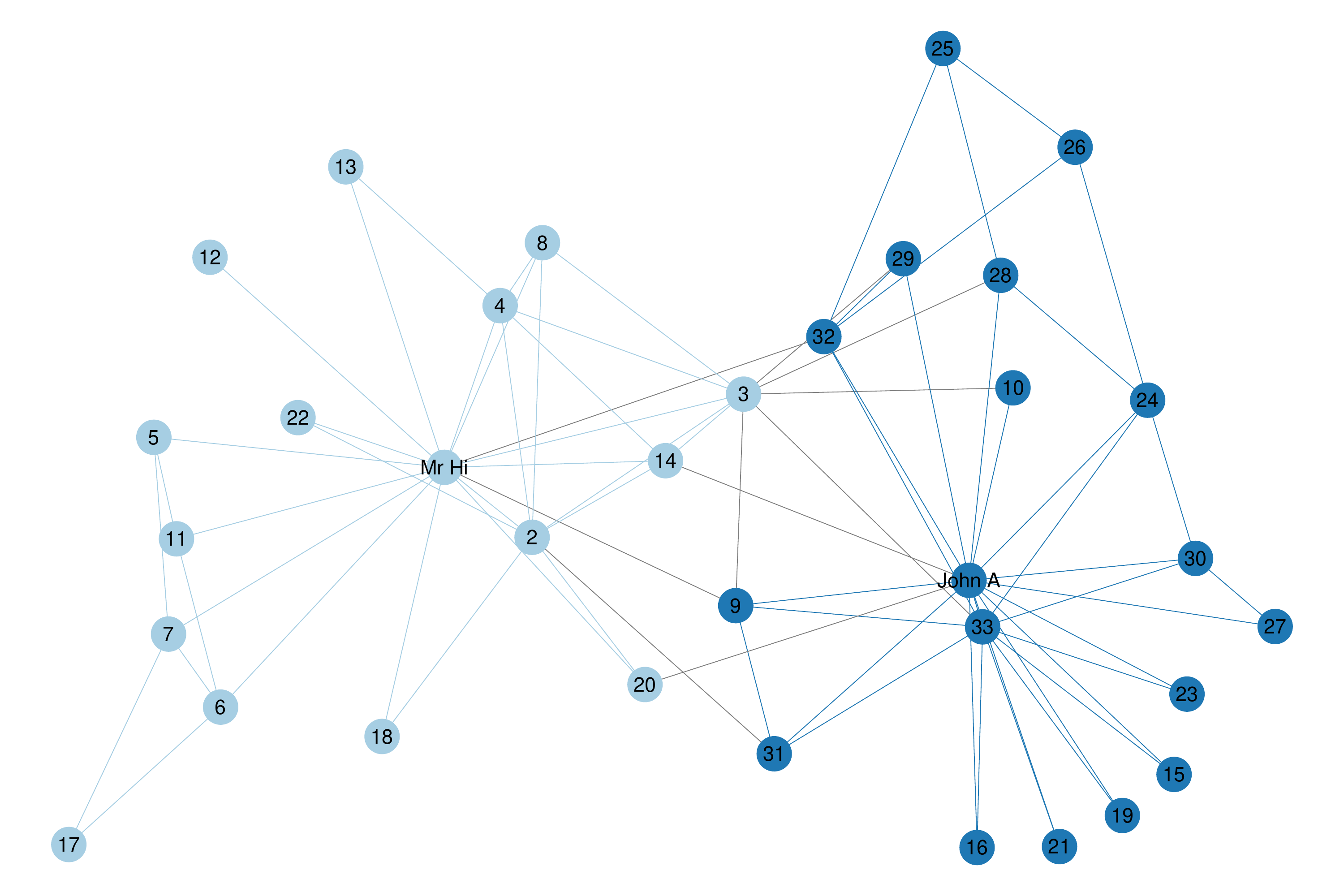}
\begin{tabular}{ccccccccc}
\multicolumn{9}{l}{binary network}\\
\hline
G-N & Louvain & SBM & \multicolumn{6}{c}{Density-based clustering}  \\\hline 
	&		  &		& degree & loc. density & betw. &&& \\ \hline
 0.58 &0.59 & 0.01& 1 & 0.36 & 1 &&&\\
%\# clusters &   5 &  4  & 3& 2 &   18 & 2   &&& \\
\hline
%\multicolumn{7}{l}{}\\
\multicolumn{9}{l}{}\\
\multicolumn{9}{l}{weighted network}\\
\hline
G-N & Louvain & SBM & \multicolumn{6}{c}{Density-based clustering}  \\
	& 		  &     & \multicolumn{3}{c}{OR option}  & \multicolumn{3}{c}{AND option}\\\hline 
	&		  &		& degree & loc. density & betw. & degree & loc. density & betw.\\\hline
 0.56  & 0.69 	  & 0.01& 1 	 & 0.44      	 & 0.85 	  & 0.61   & 0.36 		   & 0.42\\
%\# clusters & - &  4  &   2& 2 &   12 & 3  & 6 & 25 & 10\\
\hline
\end{tabular}
\caption{Zachary Karate Club network with true communities marked with different colours. Below, NMI results of different community detection methods.}\label{tab:karate}
\end{figure}

In agreement with these considerations, in the binary setting, an essentially perfect agreement is found between the two factions and the density-based partition detected with both degree and betweenness as node-wise measures. Conversely, since local density accounts for the maximum number of ties each actor can set in its neighbourhood, it results in depowering the leaders of star-based community, thus proving not to be adequate as a node-wise measure to recover the true factions. %In fact, it is worth noting that $10$ over the $34$ actors are considered of ambiguous membership. 
A slightly better performance arises from the application of both Louvain and GN methods, while SBM cannot reconstruct the benchmark factions. Note, however, that none of the competitors is designed to detect hub headed communities.  See Figure \ref{tab:karate}.

When considering the weighted network with the OR option, the two factions are again perfectly recovered with the degree used as a node-wise measure. Betwenness overall makes a remarkable job as well, althought it identifies three community instead of two. Accounting for the link weights, in fact, allows to distinguish a new leader beyond Mr Hi and John A, namely actor 32, with a high prominent bridging role. The inadequacy of local density to find clusters arising from a leadership is confirmed also in the weighted setting.  
The AND option gives rise, by construction, to a larger number of homogeneous clusters, with the highest density ones still led by John A and Mr Hi. For this reason the NMI stands at decreased values. Disregarding the employed density, in general, the presence of more peripheral actors is enhanced, as with the AND option individual connections are accounted for in clustering formation, rather than the leader influence. In fact, despite the true cluster membership, driven by a forced choice of each actor to line up with one of the leaders, data show that the relationships among the peripheral actors are generally stronger than the ones they have with the leaders. 

%While in principle applicable to weighted networks, the Girvan-Newman method interprets weights as distances instead of connection strengths. For this reason we do not report results for such method in the weighted setting. 
The Louvain method produces improved results with respect to the binary case, while SBM and GN stand at about the same level than in the binary counterpart. 

\paragraph{Les Mis\'erables character network}

\begin{figure}[t]
\centering
\includegraphics[scale=.4]{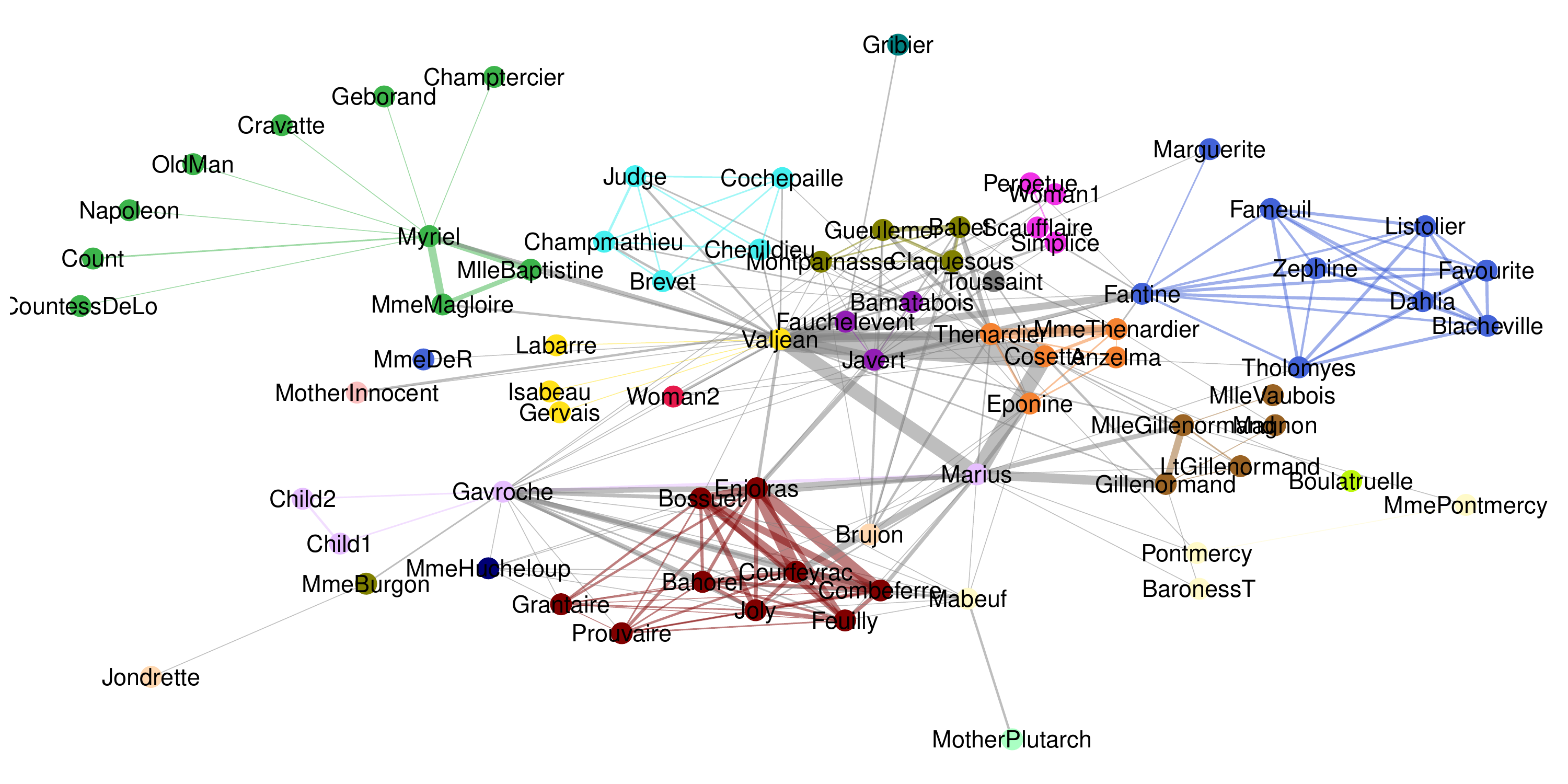}
\begin{tabular}{ccccccccc}
\multicolumn{9}{l}{binary network}\\
\hline
G-N & Louvain & SBM & \multicolumn{6}{c}{Density-based clustering}  \\\hline 
	&		  &		& degree & loc. density & betw. &&& \\ \hline
0.76&  0.63   & 0.54& 0      & 0.76         & 0     &&&\\
%\# clusters &   5 &  4  & 3& 2 &   18 & 2   &&& \\
\hline
%\multicolumn{7}{l}{}\\
\multicolumn{9}{l}{}\\
\multicolumn{9}{l}{weighted network}\\
\hline
G-N & Louvain & SBM & \multicolumn{6}{c}{Density-based clustering}  \\
	& 		  &     & \multicolumn{3}{c}{OR option}  & \multicolumn{3}{c}{AND option}\\\hline 
	&		  &		& degree & loc. density & betw. & degree & loc. density & betw.\\\hline
 0.35 	  & 0.63& 0.59 	 & 0.48      	 & 0.43 	  & 0.61   & 0.76 		   & 0.78 & 0.78\\
%\# clusters & - &  4  &   2& 2 &   12 & 3  & 6 & 25 & 10\\
\hline
\end{tabular}
\caption{Les Mis\'erables character network. Cf. Figure \ref{tab:karate}.}\label{tab:lesmis}
\end{figure}

This popular network describes the interactions between 77 characters of the Victor Hugo's novel Les Mis\'erables \citep{Knuth1993}. The network is in principle weighted, with edge strength set to the number of co-appearance of characters in one or more scenes of the novel. Like in the previous example, we also analyse its binary version. 
With the aim of an objective evaluation, we pursue the assignment of a ground truth membership by associating each character to the book of his/her early appearance. This eventually results in 20 small communities having an assorted attachment mechanism, with some communities formed around more relevant characters and other more cohesive communities (Figure \ref{tab:lesmis}). 
The partition provides an overall fair summary of the novel plot, yet we shall account with some limitations. Beyond three ambiguous references to unnamed actors, the cluster membership of a few main characters should rather have overlapping nature. Hence, results evaluation requires some further insights beyond the mere observation of the NMI values.

Neglecting the strenght of relationships, has little impact on the minor characters, generally claiming a small number of weak interactions. Thus, in the binary version of the network, cohesive communities are anyway easily detected, whereby GN, Louvain and modal clustering based on local density stand out from the other methods at high values of accuracy. Conversely, the loss of information on the weights affects the classification of the main characters, all being connected to each other, yet with a different extent which pinpoints their role. Hence, modal clustering with degree and betweenness identifies in the binary network just one community, built around the protagonist Jean Valjean. 

When the weight strength is accounted for, modal clustering works remarkably with the AND option, which tends to inflate the segmentation and highlights minor groups. The OR option underperforms the AND version compared to the true labels, yet results are anyway highly interpretable. The use of both degree and betwenness gives rise to 6 clusters. In the former case most communities are built around one main character, whereas when betwenness centrality kicks in, being its values larger for those characters having a protagonist role in multiple books, modal clustering is able to isolate all the main characters in just one group (that is the ``main plot'' cluster) together with other 5 smaller sized groups of actors whose story is standalone within the whole plot.

\paragraph{US politics books co-purchasing network}
\begin{figure}[t]
\begin{center}
\hspace{-2.5cm}\begin{minipage}{.45\textwidth}\includegraphics[scale=.3]{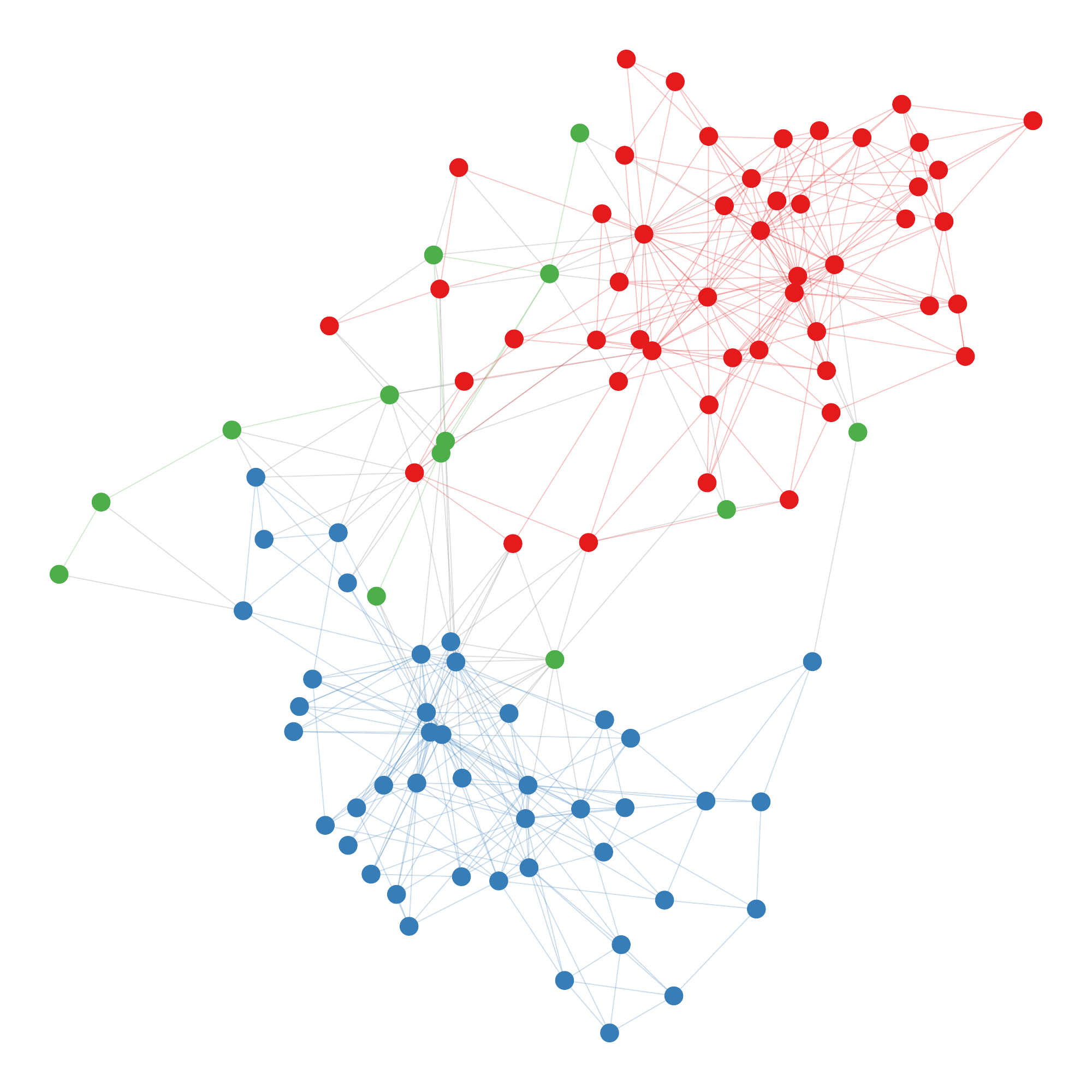}
\end{minipage}
\hspace{-0.5cm}
\begin{minipage}{.45\textwidth}
\begin{tabular}{cccccc}
%\multicolumn{9}{l}{binary network}\\
\hline
G-N & Louvain & SBM & \multicolumn{3}{c}{Density-based clustering}  \\\hline 
	&		  &		& degree & loc. density & betw.  \\ \hline
 0.56 &0.51 & 0.45& 0.60 & 0.31 & 0.07 \\
%\# clusters &   5 &  4  & 3& 2 &   18 & 2   &&& \\
\hline
\end{tabular}
\end{minipage}
\caption{US politics books co-purchasing network. Cf. Figure \ref{tab:karate}.}\label{tab:polbooks}
%
%
%\begin{table}
%\begin{tabular}{lc}
%Method & NMI\\ 
%\hline
%Density-based clustering (degree) & 0.6 \\  
%Density-based clustering (local density) & 0.31 \\
%Density-based clustering (betweenness) & 0.07 \\
%Girvan-Newman & 0.56 \\
%Louvain &  0.51 \\
%SBM &  0.45 \\
%\hline
%\end{tabular}
%\label{tab:polbooks}
%\caption{NMI values for the community detection methods applied to the US politics books co-purchasing network.}
%\end{table}
\end{center}
\end{figure}

As a further example of star-shaped communities in networks, the US politics books co-purchasing data\footnote{\url{http://www.orgnet.com/}} include 105 books about US politics published around the presidential election in 2004 and sold online at Amazon.com. The 441 ties between them represent co-purchasing of books by the same buyers. %The network was built by Valdis Krebs and published on his web site (\url{http://www.orgnet.com/}). 
Community membership is given by the book political alignment: liberal, neutral, or conservative. Within communities there exists a slighlty centralized organization of links, especially among liberal and conservative thinkings, with bestsellers representing high-density nodes, often bought in bundle with a variety of less popular other books.

Results (Figure \ref{tab:polbooks}) reflect such behaviour, as the density-based partition built on degree centrality outperforms both the other centrality measures and the competitors. 
While the latter tend to oversegment the network yet achieving acceptable results, the former result not appropriate to describe the community configuration. 
Without exception, methods are not able to identify the least characterized neutral books. 

\paragraph{Email-EU-core network}
\begin{figure}[t]
\begin{center}
%\hspace{-2.5cm}\begin{minipage}{.45\textwidth}
%\vspace{-0.2cm}
%\hspace{-2cm}
\includegraphics[scale=.4]{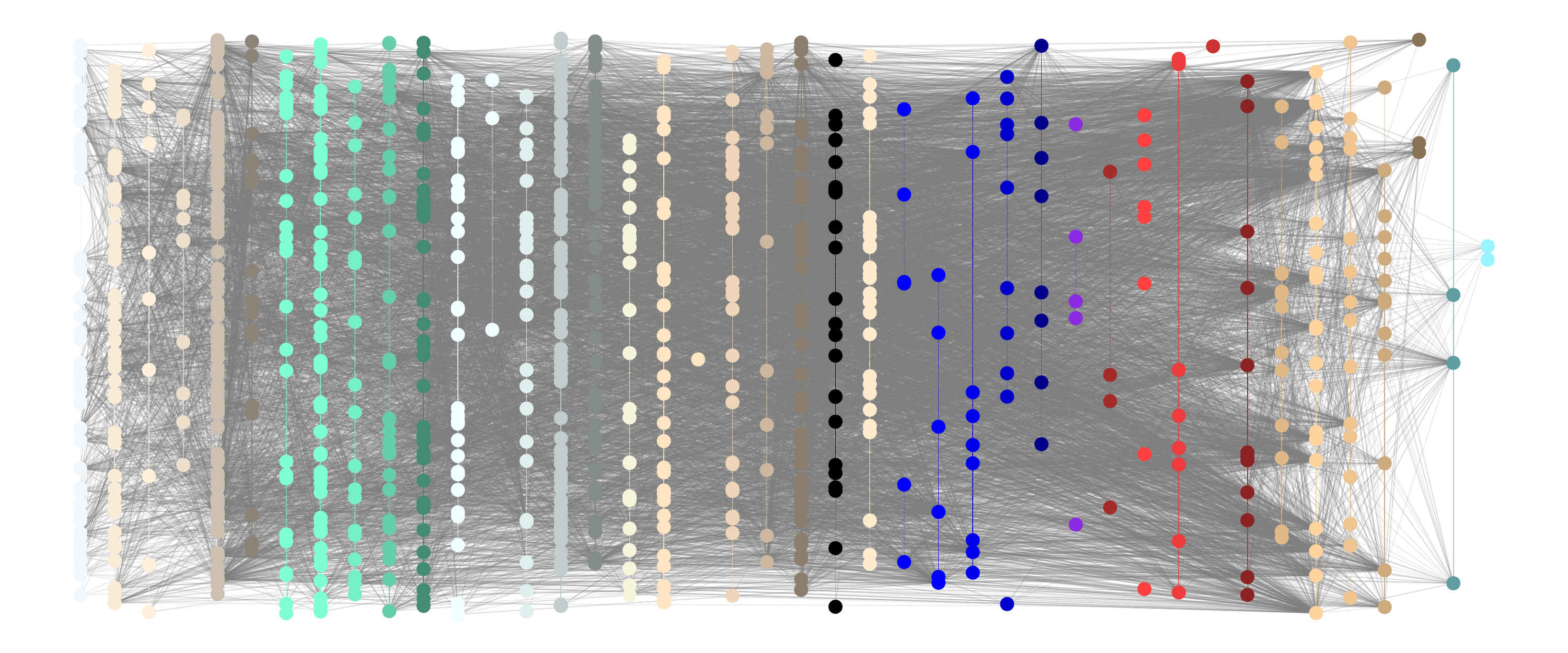}
%\end{minipage}
%\hspace{-0.5cm}
%\begin{minipage}{.45\textwidth}
\begin{tabular}{cccccc}
%\multicolumn{9}{l}{binary network}\\
\hline
G-N & Louvain & SBM & \multicolumn{3}{c}{Density-based clustering}  \\\hline 
	&		  &		& degree & loc. density & betw.  \\ \hline
 0.56 &0.53 & - & 0.26 & 0.58 & 0.26 \\
%\# clusters &   5 &  4  & 3& 2 &   18 & 2   &&& \\
\hline
\end{tabular}
%\end{minipage}
\caption{Email-EU-core network network. Cf. Figure \ref{tab:karate}.}\label{tab:mail}
\end{center}
%\begin{center}
%\begin{table}
%\begin{tabular}{ l c }
%Method & NMI\\ 
%\hline
%Density-based clustering (degree) & 0.26 \\  
%Density-based clustering  (local density) & 0.58 \\
%Density-based clustering  (betwenness) & 0.26 \\
%Girvan-Newman & 0.56 \\
%Louvain &  0.53\\
%SBM &  \\
%\hline
%\end{tabular}
%\label{tab:mail}
%\caption{NMI values for the community detection methods applied to the Email-EU-core network.}
%\end{table}
%\end{center} of the total amount of links
\end{figure}

The Email-EU-core network \citep{Leskovec2007, Hao2017} describes the email exchange between the members of 42 departments of an European research institution. The network is regarded to as undirected by setting an edge whenever there has been at least one either outgoing or incoming email between two members. True clusters are the Departments of affiliation. Distribution of actors among Departments is rather unbalanced, ranging  from 1 to 107 individuals. Since the network includes a few isolated nodes, we focus on the giant component only, consisting of 986 individuals (98\% of the total) connected by 25552 ties (99.9\%). %In fact, while our method is able to handle disconnected networks, the R routines implementing competing methods cannot do the same.  
%While edge weights are not available, the links represent the delivery of either incoming or outgoing emails    

%was generated using email data from a large European research institution in the study of \citet{Leskovec2007} and \citet{Hao2017}. The dataset contains all incoming and outgoing email between members of the research institution. We disregard direction and look at the network as undirected. Edge weights are not available. %There are a total of 1005 individuals connected by 25571 email linkages. 

%As a "ground-truth" community memberships of the nodes, we consider the 42 departments at the research institute to which the email senders belong. The distribution of nodes within communities is quite unbalanced, ranging  from 1 individual in 2 departments up to 107 individuals in the most populated department. %This happens because the e-mails in the dataset only represent communication between institution members (the core), diregarding incoming and outgoing messages outside of the research institution.

The network is far more complex than the ones examined above. While of difficult inspection, Figure \ref{tab:mail} shows that the community configuration is hardly caught by the link description. Research collaborations, indeed, are possibly conducted also by email, and likely not to be limited to the members of a Department. Additionally, there is little evidence of some attachment mechanism guided by the presence of prominent individuals in terms of their degree; also due to the unavailability of weights, conversely, it is likely to expect quite a homogeneous distribution of links within each Department and possible clusters not built around some leaders. 

Results confirm the expectations, as local density is the only centrality measure able to catch the gross community structure via density-based clustering. GN and Lovain method stand on about the same level of accuracy of classification. The application of SBM is computationally unfeasible on this network, due to an inner limitation of the R routines included in package \texttt{sbm}, which requires the joint estimation of models for any number of communities and the subsequent selection of the best models. Hence, networks with a large number of clusters, like in this case, run into a memory error. %and are not suitable to be considered.
%In fact, the transitivity within departments is quite high. The consequence is that using degree or betweenness as node-wise density measure is not suitable to detect communites. In the case of very high intra-community transitivity, it is worth to use some node-wise measures that account for the embeddedness of nodes within their respective neighborhood,  such as local or neighbor density for each node. In table \ref{tab:mail}, we report the NMI values obtained by comparing the detected community structures with the department affiliation. Modal clustering approach considering neighbor density for each node produce slightly better results than Girvan-Newman. }

\paragraph{American college football network}

\begin{figure}[t]
\begin{center}
\hspace*{-2.8cm}
\begin{minipage}{.43\textwidth}
\vspace{-0.2cm}
%\hspace{-2cm}
\includegraphics[width=7cm, height=7cm]{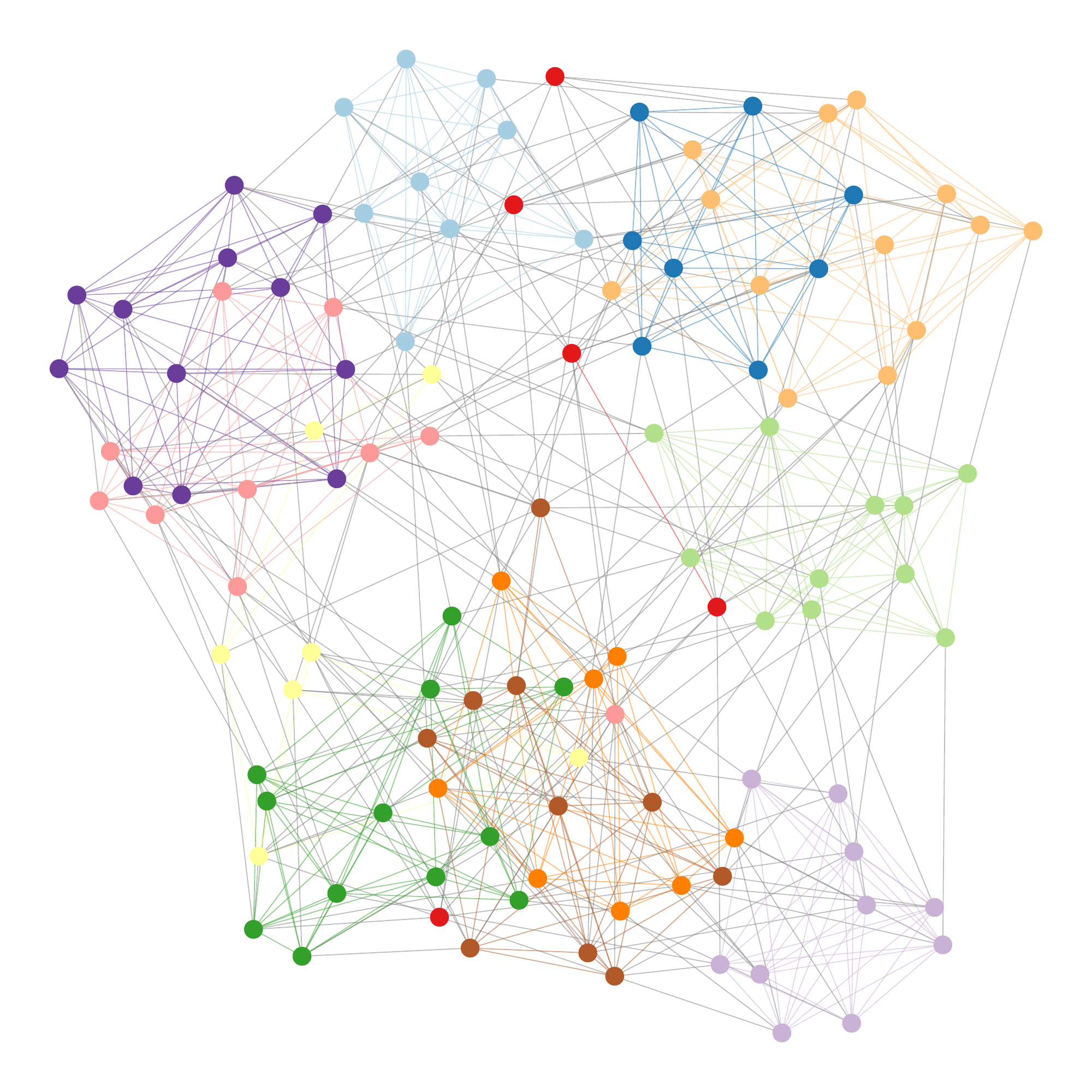}
\end{minipage}
%\hspace{-0.5cm}
\begin{minipage}{.45\textwidth}
\begin{tabular}{cccccc}
%\multicolumn{9}{l}{binary network}\\
\hline
GN & Louvain & SBM & \multicolumn{3}{c}{Density-based clustering}  \\\hline 
	&		  &		& degree & loc.dens. & betw.  \\ \hline
 0.88 &0.89 & 0.89 & 0.33 & 0.55 & 0.13 \\
%\# clusters &   5 &  4  & 3& 2 &   18 & 2   &&& \\
\hline
\end{tabular}
\end{minipage}
\caption{American college football network. Cf. Figure \ref{tab:karate}.}\label{tab:foot}
\end{center}
\end{figure}

The American college football network, described by \citet{Girvan7821}, represents the schedule of Division I American football games for the 2000 season. Nodes represent teams and ties between two teams represent regular-season games they dispute. 
The 115 teams are divided into 12 conferences, representing the benchmark community memberships. In most conferences, inner games are more frequent than games with external teams, with an average of about seven intraconference games and four interconference games in the reference season \citep[][p.7824]{Girvan7821}.
The example is here explored to show the inadequacy of density-based community detection in the lack of leadership. Games configuration, indeed, leads to a  grid-like organization of links within communities. In this situation the competing methods are able to recover accurately the community structure, while our proposal fails by setting either degree or betwenness as node-wise measure. Local density in this case, allows just for a slight improvement. See Figure \ref{tab:foot} for details. 

%\begin{center}
%\begin{table}
%\begin{tabular}{ l c }
%Method & NMI\\ 
%\hline
%Density-based clustering (degree) & 0.26 \\  
%Density-based clustering  (local density) & 0.58 \\
%Density-based clustering  (betwenness) & 0.26 \\
%Girvan-Newman & 0.56 \\
%Louvain &  0.53\\
%SBM &  \\
%\hline
%\end{tabular}
%\label{tab:mail}
%\caption{NMI values for the community detection methods applied to the Email-EU-core network.}
%\end{table}
%\end{center}

\subsection{Finding clusters within the community of Italian academic statisticians}

The aim of the case study here considered is to characterise the scientific community of the Italian academic scholars in Statistics and related fields, via the identification of the clusters formed on the basis of the relationships between them, possibly of different nature and strength, and of the leading aggregation mechanism. This can be useful, for instance, for the creation of new projects and synergies, or more generally, to understand who are, within the community, the leading actors with respect to specific topics. 

The main hypothesis underlying the data collection is that, to characterise a researcher within the community, we broadly answer to the questions:   
\emph{Where does he/she work? What is his/her macro-area of research? Who does he/she work with? What does he/she work on?}
As a consequence, we have built a weighted network having in principle a multiplex structure, divided in four layers associated with the questions above:
(1) %\emph{affiliation}, i.e.
affiliation adjacency matrix (AFF): two actors are connected when they share the same university department affiliation;
(2) %\emph{macro-area of research}, i.e
macro-sector adjacency matrix (MS): two actors are connected when they belong to the same macro-sector, within the area of Statistics and related fields, and as defined by the Italian Ministry of Education, Universities, and Research MIUR (statistics,  economic statistics, demography and social statistics, mathematical methods for economy, actuarial and financial sciences)
(3) %\emph{co-authorship}, i.e.
co-authorship network (PUBS): two actors are connected with a link weighted as the number of publications they co-authored;
(4) %\emph{research topics}, i.e.
common keywords adjacency matrix (KW): two actors are connected with a link weighted as the number of common keywords in their publications.

Data have been collected in November 2019 and refer to 1160 among professors and researchers of the academic community of statisticians, as recorded by 
the MIUR database\footnote{\url{http://cercauniversita.cineca.it}.} where information about the university affiliation and the scientific macro-sector have been drawn. Information about the publications and the keywords have been extracted from the ISI-WoS database\footnote{\url{https://apps.webofknowledge.com} }.
Handling the latter one has been troublesome, due to an awkward operation of author matching, especially in the case of homonymy or when a researcher has changed his affiliation at some time and the WoS database does not recognise it. 
In fact, we shall live with the likely, hopefully not relevant, distortion in the assessment of both the publications and the inherent keywords. 

A summarising description of the single layers is provided in Table \ref{tab:stat}. All networks at the individual layers are composed of the 1160 nodes representing the members of the scientific community under study. Given the exclusivity of the affiliation, the associated network is composed by as many components as the number of observed University departments (namely, 194), within which every actor is connected with all the other actors. The number of researchers within departments is pretty heterogeneous, ranging from 1 to 54. 
A similar behaviour is observed in the network associated to the macro-sector, where each actor is connected with all other researchers in the same macro-sector. The number of connected components in this layer is equal to the number of considered macro-sectors and these components have diverse sizes (both the statistics and mathematical methods for economy, actuarial and financial sciences areas count more that 400 researchers, while each of the two further sectors count about 150 researchers).
The co-authorship layer represents an updated, enriched version of one of the databases employed by \citet{DeStefano_etal:13}. We observe 255 isolated researchers, either because they have not published on ISI journals, or because their publications have never been co-authored by any other Italian academic statistician currently on the MIUR list. %(these include, for instance, researchers working only with non-statisticians, with foreign researchers, with PhD students or Post-doc, or with researchers that were retired at the time of data downloading).
Their publications have been in any case considered to extract the keywords for the fourth layer of the network,
where the number of isolated researchers reduces to 109.%, which have then excluded from the subsequent analyses. %We shall assume that these latter statisticians are inactive.
%Note that the co-authorship network layer represents an updated, enriched version of one of the databases employed by \citet{DeStefano_etal:13} for an analysis of scientific collaboration in the same scientific community.

\begin{table}
\caption{Italian academic statisticians network: descriptive statistics for the individual layer networks (AFF - Department affiliation, MS - Macro-sector, CA - Coauthorship network, KW - common keywords) and overall weighted networks.}\label{tab:stat}
\begin{tabular}{llccccc}
\hline					 & AFF 		    & MS 	   	   & PUBS  			& KW  & Overall\\ 
\# of isolated nodes  	 & 	67		    &   0		   & 255			& 109 & 0\\
\# of components 		 &  194 		& 	4		   & 292			& 110 & 1\\
%\# of edges &&&&&\\
Network Density 		 & 0.014    & 0.308	& 	0.002		& 0.523 & 0.734\\
Global transitivity	 	 &  1		& 	1	& 	0.306		& 0.820 & 0.833\\
Degree centralization	 & 		0.032				& 		0.082	   & 			0.016 & 0.346 	& 		0.240\\
\hline
\end{tabular}
\end{table}

In order to aggregate the four layers into a single, weighted network, we have first normalised the edge weights, measured on different scales, depending on the represented relationship.
In principle, there are many procedures to choose among for the purpose.% \citep[e.g.][]{VOROS2017}. 
We opt for the simple idea of dividing each weight by the sum of weights within the layer. 
Then, stemming from the four normalised networks, we have built the associated \emph{overlapping} network \citep{Battiston_etal:14} by simply summing up the edge weights associated to the same actor across different layers. 
%\textcolor{red}{While some information is surely lost when building the overlapping network, a stronger relationship is yet exhibited between actors which are connected at different layers.  (Toglierei)}
%Note that this choice results in weighting layers inversely for their sparsity, hence the strength of the links in the overlapping network is largely governed by the sparsest co-authorship layer.
%\textcolor{red}{The resulting overall network is represented by the $\mathbf{W}$ adjacency matrix where its element is $w_{ij}= \sum_{k \in \matchal{K}} \frac{1}{l^k_{ij}}$, where $l^k_{ij}$ is the normalize edge weight between the nodes $i$ and $j$ at layer $k$ in the $\matchal{K}$ of layers. Then, the overall weighted network edges represent the total intensity of the relations between the $i$-th and $j$-th nodes, built by summing up the edge weights associated to the same actor across different layers \citep{Battiston_etal:14}. 
%The final overall network is composed of 1051 connected nodes and is relatively dense (0.734) and cohesive (transitivity equals to 0.833). Considering the weights,
The overall network is relatively dense and cohesive with no isolates since all nodes are comprised in the unique component (see Table \ref{tab:stat}).
The strength of the links in the overall network is largely governed by the sparsest co-authorship layer because of the used weighting system.

\begin{figure}[t]
\centering
\begin{tabular}{llllll}
\hline
\multicolumn{2}{c}{Louvain} & \multicolumn{4}{c}{Density-based clustering}\\
&&\multicolumn{2}{c}{Option OR} & \multicolumn{2}{c}{Option AND}\\
\hline
size  &45.69 (22.90) & size & 7.56 (7.97) & size & 2.11 (1.42)\\
\hline
modularity &\hspace{.15cm}0.42 & modularity & \hspace{.15cm}0.24 &modularity & \hspace{.0cm}0.18\\
\hline
 \multicolumn{2}{c}{\includegraphics[scale=0.23]{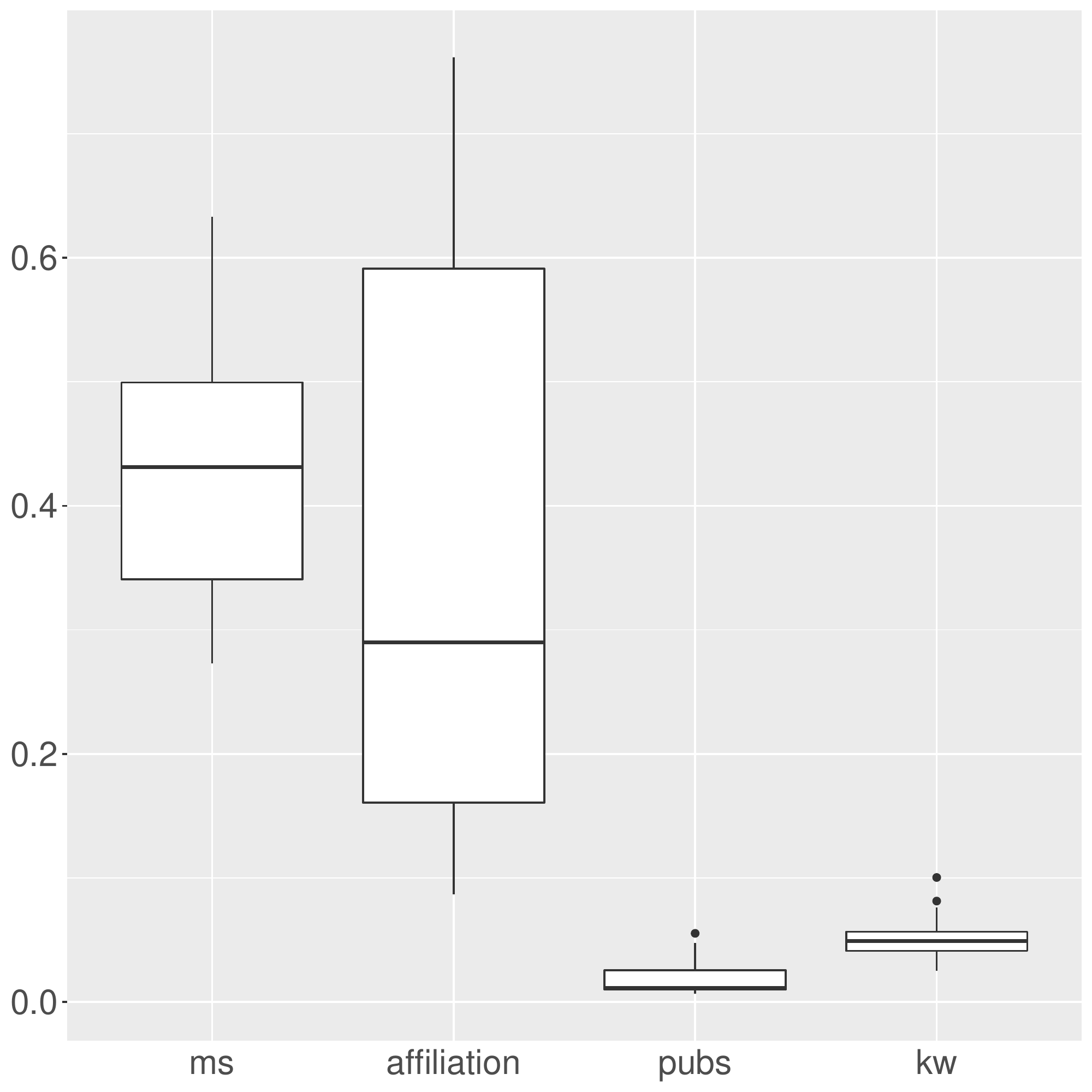}}&
 \multicolumn{2}{c}{\includegraphics[scale=0.23]{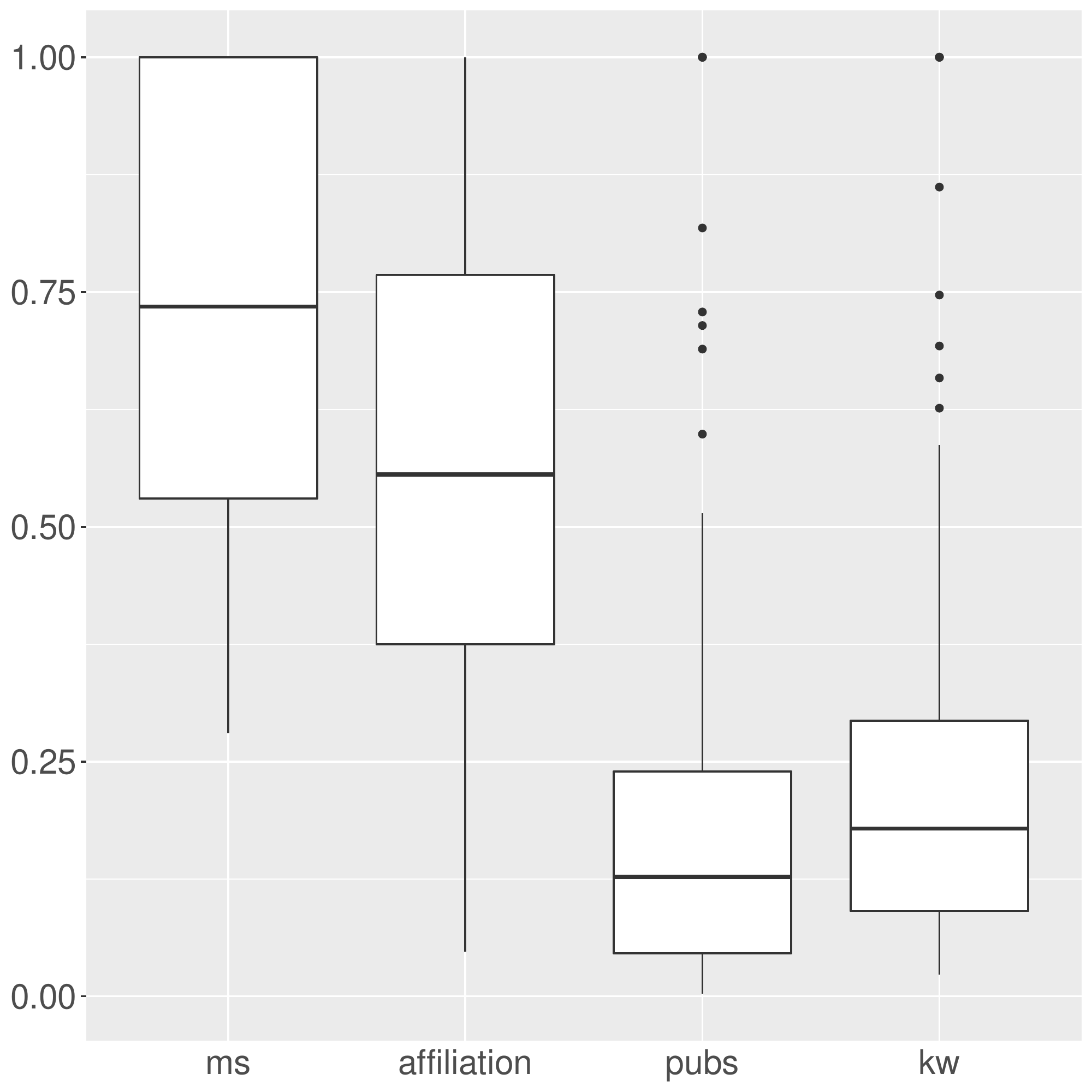}}&
 \multicolumn{2}{c}{\includegraphics[scale=0.23]{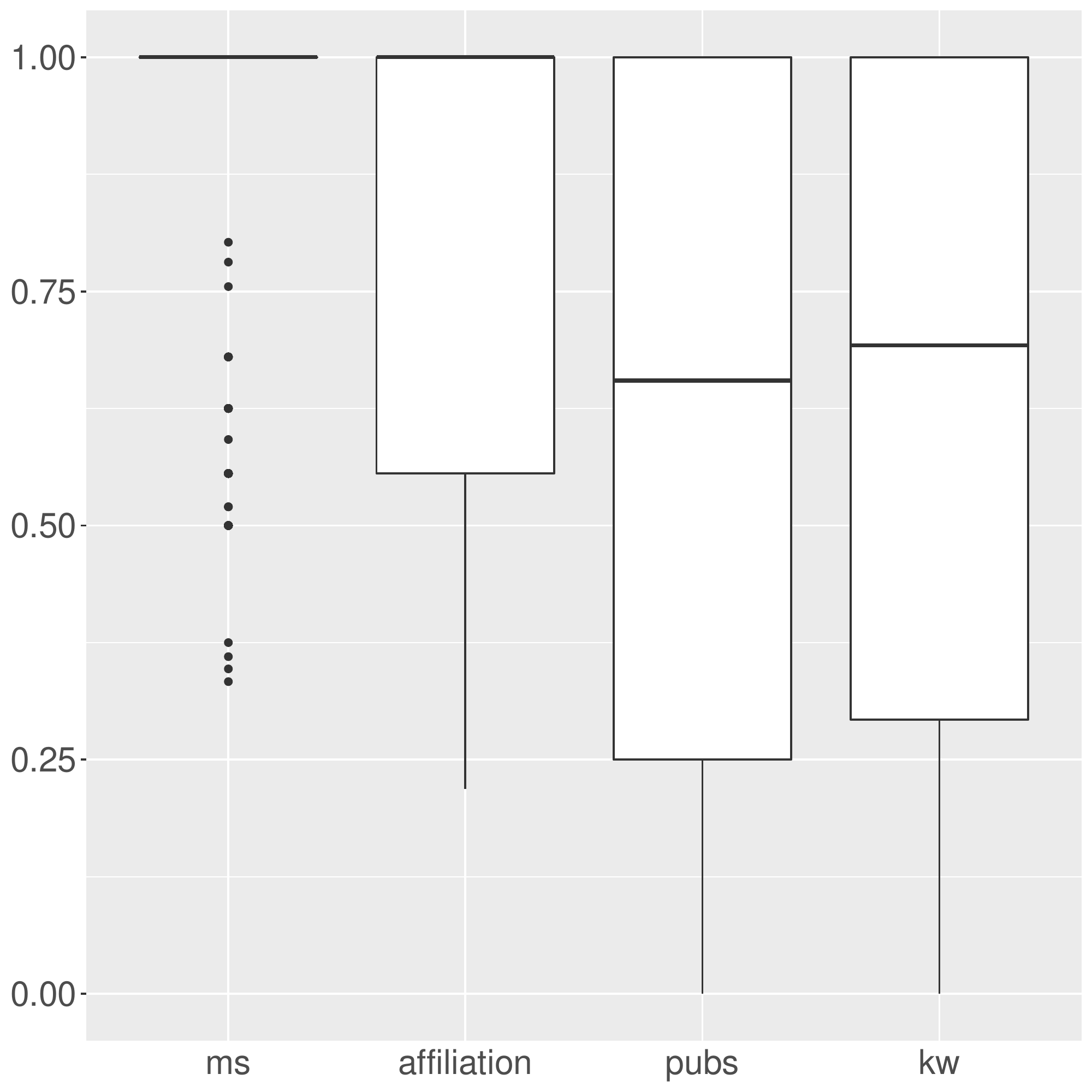}}
\end{tabular}
\caption{Italian community of statisticians. Top panel: average size of clusters (and standard deviation) found via the Louvain method and both the options OR/AND of the  density-based method and modularity. The boxplots display the homogeneity of actors across clusters with respect to the considered relationships.}\label{fig:boxplot}
\end{figure}

Among the community detection methods, we have been able to run the Louvain method only, whereas the application of both GN and SBM has turned out computationally unfeasible. It is worth reporting, however, that we run SBM up to the maximum number of communities allowed by our computational resources, i.e. 64. The detected partition is unarguably suboptimal, with one community gathering the $2/3$ of the actors but the remaining 63 clusters do not differ much from the ones obtained with our procedure.   
The Louvain algorithm identifies $23$ communities, of size ranging from 13 to 102 researchers. Cluster homogeneity with respect to the scientific macro-sector and the affiliation has been evaluated via the complement to one of the Gini index. As for the publications, for each researcher, the proportion of works coauthored by members of the same cluster has been evaluated and the cluster average used as a summarising measure of cluster homogeneity. The same index has been computed for the keywords. To look at the assortative mixing within the detected communities and find if actors within clusters tend to exhibit dense connections among them rather than with actors in different clusters, modularity of the clusters has been also evaluated.
Results are summarized in Figure \ref{fig:boxplot}. Due to the large size of the detected clusters, clusters are somewhat homogeneous for the scientific sector and the affiliation only, %\textcolor{red}{(these relations alone are not meaningful in terms of a real collaboration between scholars)}, 
while communities are scarcely associated to co-authorship and research topics. While, by construction, modularity of the Louvain-based partition is maximised, it does not show a remarkably high value. To this respect, it is worth noting that even if the detected partition corresponds to the global maximum of the modularity, in scientific applications this solution is not guaranteed to be more meaningful than the ones obtained by local maxima \citep{Good_2010}. %Furthermore, the resolution limit affecting modularity optimisation algorithms prevents to find smaller clusters that can be meaningful in terms of scientific collaboration. 
Furthermore, results are affected by the so-called resolution limit for which small, plausible communities cannot be identified if the network is large and heterogeneous clusters tend to be formed \citep{FortunatoResolution}. %This leads to some questionable results in some application domains. 

Modal clustering has been run building on the degree of the actors, as it appears the most sensible and easiest to interpret choice in such a complex application. Both the options ``OR'' and ``AND'' have been run. Summarising results in terms of size of clusters, modularity, and homogeneity with respect to the considered relationships are reported in Figure \ref{fig:boxplot}. The OR option identifies 139 groups of size ranging from 4 to 49 scholars.
Some heterogeneity with respect to the considered relationships is unavoidable, but clusters are far more homogeneous than those identified by the Louvain method. In fact, while actors working either together or on similar topics tend to be aggregated into the same cluster, the same membership is often shared by other researchers. In this case, cluster aggregation is mostly driven by the attraction hold by a few leaders towards minor actors which often exhibit pretty diverse characteristics.   %\textcolor{red}{possiamo dire che questi leader sono full professor o cmq senior e piu importanti degli altri anche scientificamente? se non si puo fa niente}
Conversely, option AND gives rise to a very sensible partition, counting 499 clusters, overall a realistic value in the overview of the statistical community where, excluded applied interdisciplinary scholars, researchers tend to work within very small-sizes teams, and publications are generally co-authored by 2/3 researchers at most. The majority of clusters is fully homogeneous with respect to the scientific macro-sector and affiliation, and gathers researchers who are known to belong to the same research group. Note that this result is largely acknowledged to occur in social contexts, where social groups are limited in size even if social actors are embedded in relatively large networks \citep{Dunbar1992}.

%Due to a smaller cluster size than Louvain groups, modularity is lower both using the AND and the OR option. This is because the method generally accounts for a larger resolution of the detected community structure. 

%Also due to a smaller cluster size than Louvain groups, modularity is lower both using the AND and the OR option. This is because the method generally accounts for a larger resolution of the detected community structure. In scientific network applications, if the purpose is to look at the possible synergy among researchers for new projects or to favour emerging topics in the field, probably it is more appropriate to consider the AND option and hence the emergence of possibly smaller and more homogeneous clusters than those detected by means of the OR option or by the Louvain algorithm.

\begin{figure}[t]
\begin{center}
%\multirow{ 2}{*}{
\hspace*{-.5cm}\includegraphics[angle=90,width=1.1\textwidth ]{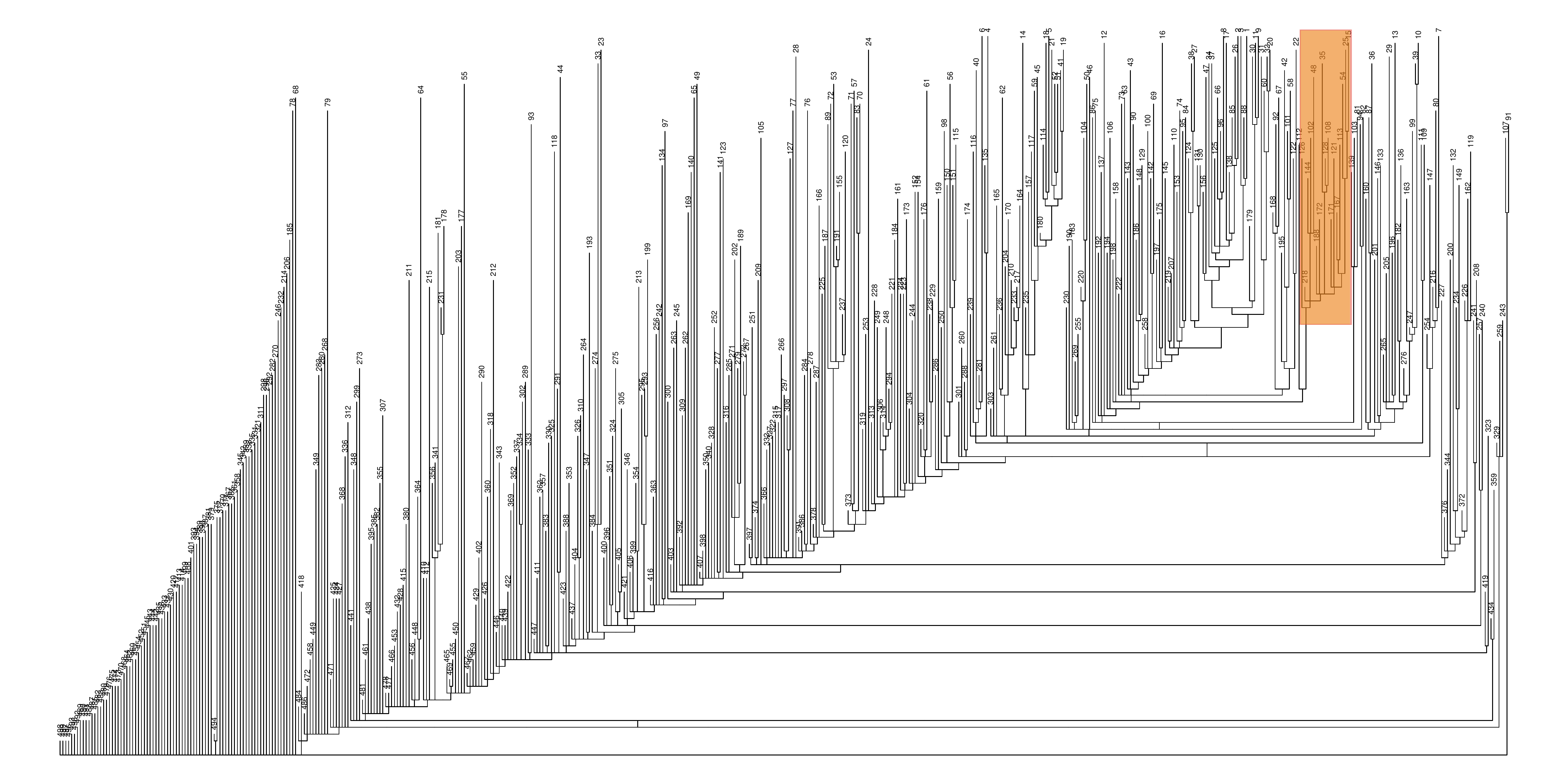}
\end{center}
\caption{Cluster tree of the Italian academic statisticians (rotated for better readability, with high density levels at the right side). An insight of the highlighted area is provided in Figure \ref{fig:zoom}.
}\label{fig:tree}
\end{figure}

\begin{figure}[t]
    \centering
    \includegraphics[scale=.13]{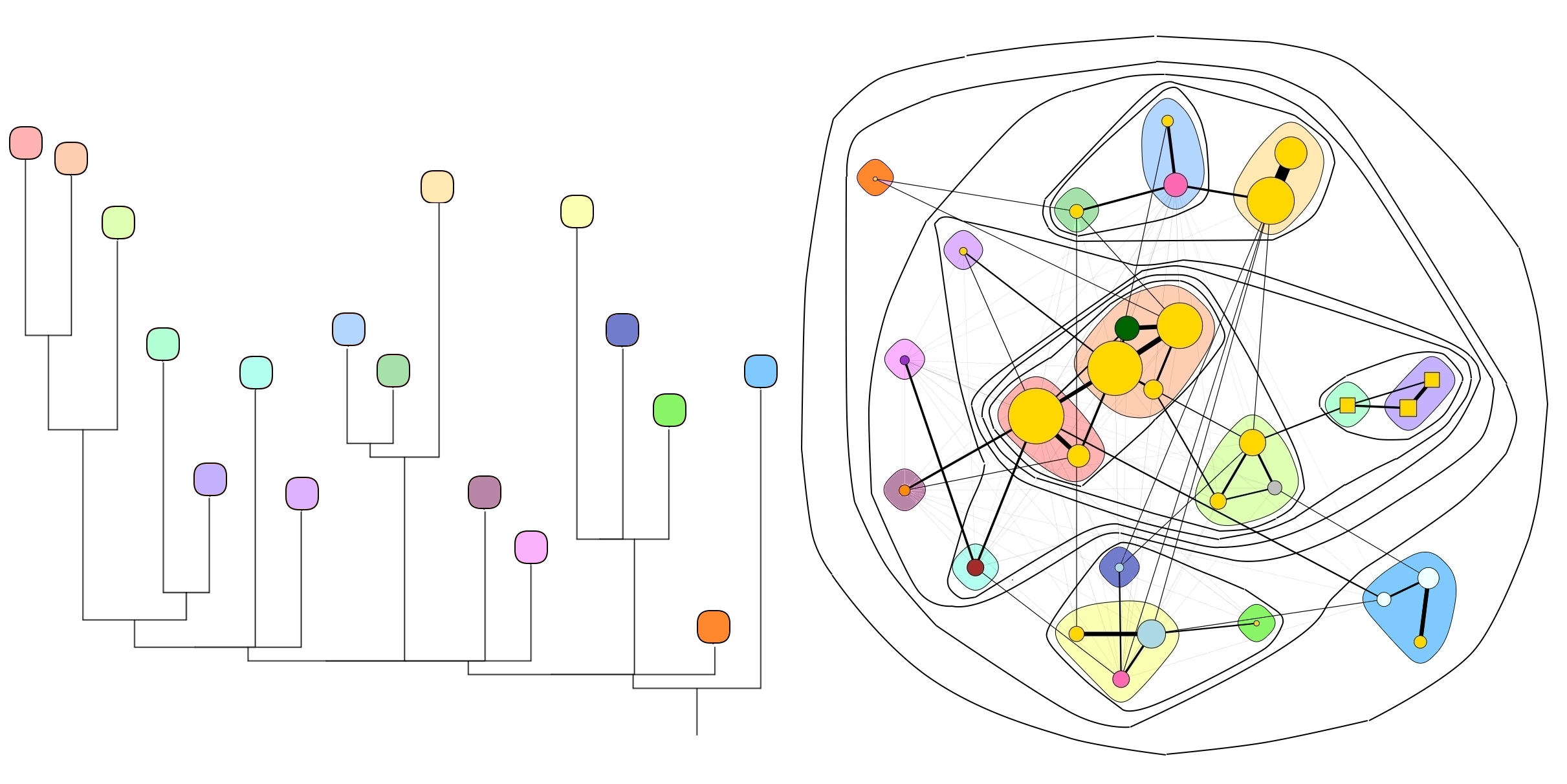}
    \caption{Detailed visualization of the subtree highlighted in Figure \ref{fig:tree} and associated subnetwork with clusters marked in different colors and superimposed the cluster aggregations at the different levels of the cluster tree. Actor size is proportional to the their density and different shapes are associated to different macro-sectors. Actor colour is associated to the affiliation. Edge width is proportional to the number of common keywords and coauthored publications.}
    \label{fig:zoom}
\end{figure}

However, the clustering is not to be interpreted solely in term of final group membership
 - a not that different partition could be trivially obtained by aggregating pairs of maximally connected actors.
In fact, the generation process of collaboration among researchers is pretty peculiar: there may be solitary researchers,
sparsely collaborating with other subjects; also, there are
researchers who mostly focus on a specific research topic, but also collaborate with different groups of people
on a variety of different areas. Both keeping these researchers separated or merging them into the same group may be a stretch.  To this aim, a relevant interpretation derives from the exploration of the cluster tree, where clusters are subsequently aggregated     
at lower levels of the hierarchy, to form larger clusters with a lower resolution (Figure \ref{fig:tree}). For the sake of interpretation, one of its branches, including 17 clusters and a total of 30 researchers, is detailed in Figure \ref{fig:zoom}, along with the associated subnetwork highlighting cluster aggregations at the different levels of the cluster tree.   
The forming leaves of the branch mostly include either researchers affiliated to the Department of Statistical Science at University
 of Padova, or scholars who have spent at that Department part of their academic career. 
Actor aggregation in clusters mostly relies on the strength of connections, hence lead by co-authorship which weights most on the overlapping network. At a lower level of the tree, cluster merging is driven by research topics, with the largest branch on the left associated to likelihood theory, and the other branches including scholars working on its more applied declinations.  
A link between  branches derives from the eclecticism of some of the researchers, working on different research topics.
The size of the tree prevents an overall interpretation, but similar traits of homogeneity can be easily identified
by picking any branch of the tree. Of course, the lower the level of aggregation of the branches, the lower the homogeneity of
the branch.    

\section{Discussion}

%\textcolor{red}{In the present paper we proposed a network clustering approach that differs from the existing methods %in the voluminous panorama of the community detection literature 
%in shifting the focus from a tie-based perspective to a node-oriented approach. Here we list the peculiarity of our proposal with respect to the the commonly used community detection algorithms: \emph{i)} we look for groups of nodes determined because of the presence high-density actors rather than because nodes are only densely and uniformly connected; \emph{ii)} the flexibility of the approach allows for the identification of clusters determined by rather diverse attachment mechanisms (e.g., leadership based communities) or according to subject specific aims accounted for diverse specification of the node-wise density function; \emph{iii)} thanks to the properties of the modal clustering, in our approach we can account for different levels of cluster resolution, via the group hierarchy provided by the cluster tree, then we are able to find even small clusters independently on the network size if this is the purpose.  (Io metterei nelle conclusioni proprio un elenco anche non esaustivo delle peculiarità del metodo ceh un po sintetizza la discussione estesa che segue)}

Due to the unsupervised nature of the problem, and to the further lack of a ground truth against which to
evaluate the quality of a partition, clustering is an ill-posed task, which cannot be performed fully automatically,
i.e. without some amount of human intervention and disregarding subject-matter considerations.

The methodology here presented makes no exception in the clustering panorama, as it both has required
during its planning and still requires the user to make a few thorny choices. 
A first choice concerns the density measure. The lack of a probabilistic notion of density at a node-wise level implies the loss, for network data, of the probabilistic framework of the original approach defined for non-relational data. Hence, the proposed procedure cannot enjoy the mathematical rigour of other well known stochastic procedures. 
On the other hand, it follows the opportunity of selecting the measure of density among a wide set of candidates which quantify connectivity or centrality roles of the actors. 
Different group structures arise according to the chosen density measure and those structures account for different aspects of subnetworks cohesiveness. %The lower the rank correlation among the competing measures, the larger the differences among the identified group structures. 
In fact, we believe that leaving unspecified this measure represents a strength of the procedure.
Depending on subject-matter considerations, this provides the procedure with the flexibility of adapting to different
notions of clusters, each of them associated with a specific selection of
the density and consistently with the intrinsic ill-posedness of the clustering problem.  

A second choice concerns the way to handle relationships of different strength. Unlike the unweighted framework,
there is no obvious way to extend modal clustering in the presence of weighted links. 
Our strategy aggregates 
strongly connected individuals at a higher density level than individuals which are weakly connected.
While this choice is consistent with the considered aggregation mechanism, based on the most prominent actors exerting
 influence over their neighbours, the actual implementation of this idea may take various forms.
 The AND option aggregates two actors with density above a threshold, when they represent their reciprocal strongest
 connection among those not examined yet. Alternatively, the condition may be loosen via the OR option, by requiring that such connection is the strongest for just one of
 the actors. A further alternative route would consist in proceeding in a block-sequential manner, aggregating several actors with density above the threshold
at a time, as long as their relationship has, at least, a given strenght. 
The possibility of choosing among these options in cluster formation allows for looking at a given network
structure from a different granularity of the  representation.  As showed in the proposed applications, the OR mechanism tends to minimize the network partition in a smaller number of internal densely connected clusters with loose connections with other clusters. On the other hand, the AND mechanism maximizes the internal homogeneity of clusters detecting a larger number of smaller groups. 
Here again the choice of the mechanism to handle weights depends on the purpose of the analysis. 
For instance in the Karate network, the choice of the OR option would reflect an interest in the big picture after collapsing the relations in the community. Conversely, in the Italian statisticians network, the choice of the AND option would entail small scale groups of actors and reflect the purpose of looking for cohesive research clusters.

Although featuring these different options of analysis, the proposed density-based procedure does not suffer from the arbitrariness matters which are typical of standard clustering procedures. While the number of clusters is determined within the procedure, the partitioning accounts for different levels of cluster resolution, via the group hierarchy provided by the cluster tree.
In this sense, the cluster tree represents a somewhat formal instrument to emulate the human cognitive system and allows for getting over the resolution limit of modularity-based methods.%, and then once more, it strengthens the complying of the cluster concept with a ‘natural’ grouping of data. 
%This is prevented by standard hierarchical clustering, where an (arbitrary) cut of the dendrogram is required to define an univocal group assignment. 
%Finally, unlike optimisation methods which suffer from the resolution limit, as the popular Louvain algorithm, we are able to find small communities even in huge networks and jointly account for larger levels of aggregation.

From a computational point of view, the algorithm requires to run $O((V+E)V)$ operations on a binary network, in addition to the ones needed to compute the density, which depend on the selected node-wise measure. While the quadratic growth discourages the use of the procedure with huge networks, we have not experienced system crashes in any of the examples run in the manuscript, and the procedure has proven its feasiblity on networks having $V$ in the order of the thausands. According to our experience, detecting communities in networks of such size is conversely prevented by the application of popular competitors like Girvan Newman and SBM.

\bibliographystyle{rss}
%\bibliography{biblio}

\end{document}